\documentclass[twocolumn,trackchanges]{aastex63}
\usepackage{graphicx}
\usepackage{hyperref}
\usepackage{epstopdf}
\usepackage{CJK}
\usepackage{verbatim}
\begin{document}
\begin{CJK}{UTF8}{gbsn}

\title{Five New Post-Main-Sequence Debris Disks with Gaseous Emission} 

\correspondingauthor{Erik Dennihy}
\email{edennihy@gemini.edu}

\author[0000-0003-2852-268X]{Erik Dennihy}
\affil{Gemini Observatory/NSF's NOIRLab, Casilla 603, La Serena, Chile}

\author[0000-0002-8808-4282]{Siyi Xu (许\CJKfamily{bsmi}偲\CJKfamily{gbsn}艺)}
\affil{Gemini Observatory/NSF's NOIRLab, 670 N. A'ohoku Place, Hilo, Hawai'i, 96720, USA}

\author[0000-0001-9372-4611]{Samuel Lai (赖民希)}
\affil{Gemini Observatory/NSF's NOIRLab, 670 N. A'ohoku Place, Hilo, Hawai'i, 96720, USA}

\author[0000-0002-8070-1901]{Amy Bonsor}
\affil{Institute of Astronomy, University of Cambridge, Madingley Road, Cambridge CB3 0HA, UK}

\author{J. C. Clemens}
\affil{University of North Carolina at Chapel Hill, Department of Physics and Astronomy, Chapel Hill, NC 27599, USA}

\author{Patrick Dufour}
\affil{D\'epartement de Physique, Universit\'e de Montr\'eal, C.P. 6128, Succ. Centre-Ville, Montr\'eal, Qu\'ebec H3C 3J7, Canada}
\affil{Institut de Recherche sur les Exoplan\'etes (iREx), Universit\'e de Montr\'eal, Montr\'eal, QC H3C 3J7, Canada}

\author[0000-0002-2761-3005]{Boris T. G\"ansicke}
\affil{Department of Physics, University of Warwick, Coventry CV4 7AL, UK}
\affil{Centre for Exoplanets and Habitability, University of Warwick, Coventry CV4 7AL, UK}

\author[0000-0002-6428-4378]{Nicola Pietro Gentile Fusillo}
\affil{European Southern Observatory, Karl-Schwarzschild-Str 2, D-85748 Garching, Germany}

\author{Fran\c{c}ois Hardy}
\affil{D\'epartement de Physique, Universit\'e de Montr\'eal, C.P. 6128, Succ. Centre-Ville, Montr\'eal, Qu\'ebec H3C 3J7, Canada}

\author[0000-0003-4145-3770]{R. J. Hegedus}
\affil{University of North Carolina at Chapel Hill, Department of Physics and Astronomy, Chapel Hill, NC 27599, USA}

\author[0000-0001-5941-2286]{J. J. Hermes}
\affil{Boston University, Department of Astronomy, Boston, MA 02215, USA}

\author{B. C. Kaiser}
\affil{University of North Carolina at Chapel Hill, Department of Physics and Astronomy, Chapel Hill, NC 27599, USA}

\author[0000-0002-5908-1488]{Markus Kissler-Patig}
\affil{European Space Agency - European Space Astronomy Centre, Camino Bajo del Castillo, s/n., 28692 Villanueva de la Ca{\~{n}}ada, Madrid, Spain}

\author[0000-0001-5854-675X]{Beth Klein}
\affil{Department of Physics and Astronomy, University of California, Los Angeles, CA 90095-1562, USA}

\author[0000-0003-1543-5405]{Christopher J. Manser}
\affil{Department of Physics, University of Warwick, Coventry CV4 7AL, UK}

\author[0000-0003-1862-2951]{Joshua S. Reding}
\affil{University of North Carolina at Chapel Hill, Department of Physics and Astronomy, Chapel Hill, NC 27599, USA}

\begin{abstract}
Observations of debris disks, the products of the collisional evolution of rocky planetesimals, can be used to trace planetary activity across a wide range of stellar types. The most common end points of stellar evolution are no exception as debris disks have been observed around several dozen white dwarf stars. But instead of planetary formation, post-main-sequence debris disks are a signpost of planetary destruction, resulting in compact debris disks from the tidal disruption of remnant planetesimals. In this work, we present the discovery of five new debris disks around white dwarf stars with gaseous debris in emission. All five systems exhibit excess infrared radiation from dusty debris, emission lines from gaseous debris, and atmospheric absorption features indicating on-going accretion of metal-rich debris. In four of the systems, we detect multiple metal species in emission, some of which occur at strengths and transitions previously unseen in debris disks around white dwarf stars. Our first year of spectroscopic follow-up hints at strong variability in the emission lines that can be studied in the future, expanding the range of phenomena these post-main-sequence debris disks exhibit.  
\vspace{1cm}
\end{abstract}

\section{Introduction}

Planetary debris is a common feature of young stars, with detections of circumstellar debris disks spanning a wide range of stellar spectral classes and properties \citep{wya08:araa46}. Observations of the dusty and gaseous components of the debris disks provide a unique view of planetary formation, as their structures and evolution reveal complex planetary formation histories \citep{hug18:araa56}. In a pleasing juxtaposition, planetary destruction results in a different class of debris disks around white dwarf stars with many of the same observable characteristics, such as excess infrared radiation from dusty debris \citep{zuc87:nat330,gra90:apj357} and emission lines from gaseous debris \citep{gan06:sci314}. For recent reviews of these observation properties, see \citet{far16:nar71} and \citet{che20:natas4}. 

Debris disks around white dwarf stars are largely agreed to be the product of the tidal disruption of a planetary body that was scattered onto a highly eccentric orbit, and passed within the tidal disruption radius of the white dwarf star \citep{jur03:apjl584,ver14:mnras445}. Unlike their main-sequence counterparts, debris disks around white dwarf stars are faint and compact with outer radii on the order of one solar radius, and often escape detection \citep{koe14:aap566,roc15:mnras449, bon17:mnras468}. Despite these challenges, observations of white dwarf debris disks provide an opportunity to learn about the evolution of debris disks, as their compact nature leads to high levels of variability on short timescales that has been observed on timescales ranging from hours, to months, to years. 

Focused studies on the dusty components of individual systems reveal stochastic increases and gradual decays in infrared fluxes, suggesting collisional dust production \citep{far18:mnras481} and disruption events \citep{xu14:apjl792,wang19:apjl886}, similar to what has been observed around main-sequence debris disks \citep{su19:aj157}. Statistical studies of the larger sample indicate this type of variability could be widespread \citep{swa19:mnras484}, albeit not as pronounced in most systems.

Some of these debris disks also host observable gaseous components in addition to dusty components. The gaseous components in white dwarf debris disks offer the chance for ground-based follow-up via moderate- and high-resolution spectroscopy of absorption and emission lines. These features exhibit a wide range of evolution, including broad profile asymmetry shifts consistent with the precession of global patterns within the disk \citep{man16:mnras455, man16:mnras462, har16:aap593, den18:apj854, cau18:apjl852, for20:apj888} and changes in strength associated with gas production and depletion \citep{wil14:mnras445}. Unfortunately, systems with observable gaseous components are rare. Discoveries of systems with gas in absorption are limited by strict inclination constraints, while white dwarf stars with gaseous components in emission are estimated to have an occurrence rate among all white dwarf stars of just 0.067 percent, with fewer than ten such systems confirmed \citep{man20:mnras493}. 

In this paper, we present the follow-up of five new white dwarf stars that host debris disks with gaseous debris in emission. All five systems show signs of accretion of metal-rich debris onto the white dwarf surface, strong infrared excesses indicative of dusty debris, and double-peaked emission lines from gaseous debris in a metal-rich circumstellar environment. Three of the systems show significant evolution of their gaseous emission features over the first year of monitoring, and others exhibit multiple strong metal species in emission, expanding the range of phenomena these systems observe. 

We begin by discussing the initial discovery of the targets and the data collected in their follow-up. Next we examine the properties of the white dwarf host star, including the confirmation of atmospheric pollution from the recent accretion of metal-rich material. We continue by presenting the excess infrared radiation for each system, which traces the dusty component of the debris disk surrounding the white dwarf star. Finally we explore the emission lines emanating from the gaseous components of the debris disks, including the different metal species observed to be in emission and early results on the variability of the emission lines. We close with a brief discussion of the new systems and their addition to the sample of white dwarf stars with observable debris disks.

\section{Data Collection}

With the announcement of hundreds of thousands of new white dwarf stars from the second \emph{Gaia} Data Release \citep{jim18:mnras480,gen19:mnras482}, several hundred new debris disk candidates have been discovered around white dwarf stars by searching for excess infrared radiation consistent with dusty debris (\citealt{reb19:mnras489}; Xu et al. 2020, submitted to ApJ). Spectroscopic follow-up of these samples is ongoing, and here we present our observations of five debris disks around white dwarf stars which show gaseous debris in emission in addition to their infrared excess. These systems were independently discovered by and are also discussed in \citealt{mel20:arxiv}. In the \emph{Gaia} white dwarf candidate catalog of \citet{gen19:mnras482}, the white dwarf stars are identified as WDJ034736.69+162409.73, WDJ061131.70$-$693102.15, WDJ064405.23$-$035206.42, WDJ162259.65+584030.90, and WDJ210034.65+212256.89. For the remainder of this work they will be referred to as WD J0347+1624, WD J0611$-$6931, WD J0644$-$0352, WD J1622+5840, and WD J2100+2122 respectively. 

These objects belong to a class of white dwarf debris disk systems that provide three different observables to study the circumstellar environment and its interaction with the white dwarf star: atmospheric pollution from the ongoing accretion of metal-rich material at the white dwarf surface, excess infrared radiation from the dusty component of the debris disks, and broad emission lines from the gaseous component of the debris disks. The study of all three of these observables requires ground- and space-based follow-up, utilizing photometric and spectroscopic techniques (e.g. \citealt{mel10:apj722}). Here we present the spectroscopic and photometric data collected for the initial characterization of the five systems discussed in this paper. 

\subsection{Spectroscopic Follow-up}

We collected broad-band and high-resolution spectroscopic data for each target to characterize the white dwarf stellar atmosphere and to search for signs of ongoing accretion from the circumstellar material. In addition, several epochs of spectroscopic follow-up were dedicated to searching for variability in the emission lines. Details including the full wavelength coverage, exposure times, and resolving power for each observation are given in the Appendix in Table \ref{tab:obsspec}. 

\emph{Gemini/GMOS}:
We used the Gemini Multi-Object Spectrographs (GMOS; \citealt{hoo04:pasp116,gim16:spie9908}) on both Gemini-N and Gemini-S to perform our initial search for emission features as part of programs GN-2019A-FT-202 and GS-2019A-FT-201. These observations were focused on detection of the calcium infrared triplet emission feature at 850\,nm and used the R400 grating with the 0.5 arcsecond slit for a resolving power of 1900. In addition, we later utilized a higher-resolution setup to confirm the emission features detected in WD J1622+5840 that included the R831 grating with the 0.5 arcsecond slit for a resolving power of 3800 as part of program GN-2019A-FT-209. The GMOS data were processed using a combination of the Gemini IRAF package and an optimal extraction routine based on the methods described in \citet{mar89:pasp101}.

\emph{VLT/X-Shooter}:
We observed four of our targets with the X-Shooter wide-band intermediate-resolution spectrograph on the Very Large Telescope \citep{ver11:aap536} in stare mode, using the 1.0 arcsecond slit aperture for the UVB arm and 0.9 arcsecond slit aperture for the VIS arm as part of programs 0103.C-0431(B), 0104.C-0107(A), and 1103.D-0763(D). The X-Shooter data for 0103.C-0431(B), 0104.C-0107(A) were reduced by the ESO quality control group, and the data for 1103.D-0763(D) were reduced using the standard procedures within the {\sc Reflex}\footnote{http://www.eso.org/sci/software/reflex/} reduction tool developed by ESO. Telluric line removal was performed on the reduced spectra using {\sc MOLECFIT} \citep{sme15:aap576, kau15:aap576}. The X-Shooter spectra serve multiple purposes, providing broad spectroscopic coverage for spectral classification, sufficient resolution for the detection of narrow absorption lines, and coverage of the emission features including the calcium infrared triplet.

\emph{Keck/HIRES}:
To probe for the narrow absorption lines that confirm the white dwarf star is accreting material, we collected high-resolution echelle spectroscopy using the blue collimator of the High Resolution Echelle Spectrometer (HIRESb; \citealt{vog94:spie2198}) on the \emph{Keck}-1 10-m telescope for three of our targets visible from the northern hemisphere. The Keck HIRES data were reduced with the MAKEE\footnote{\url{https://www.astro.caltech.edu/~tb/makee/}} package, and we produced order-merged, normalized, signal-to-noise weighted combinations in regions of interest around known transitions to search for narrow absorption features.  

\emph{SOAR/Goodman}:
The Goodman spectrograph on the SOAR 4-m telescope \citep{cle04:spie5492} was employed to confirm the presence of and search for variability in the calcium infrared triplet emission features detected in all five systems. We used the 1200l-R grating with the 1.0 arcsecond slit for a resolving power of 3000. The data were reduced using a custom set of Python-based tools and an optimal extraction routine based on the methods described in \citet{mar89:pasp101}.

\subsection{Photometric Follow-up \label{sec:photfollow}}

We collected infrared photometry for each system using ground- and space-based facilities. Below we discuss the telescopes and instruments used, including details on the data processing methods. Additional details including the the filters and exposure times used are given in the Appendix in Table \ref{tab:obsphot}.

\emph{Gemini-N/NIRI}:
\emph{J}, \emph{H}, and \emph{K}-band photometry was collected using the Near InfraRed Imager and spectrograph at Gemini-North (NIRI; \citealt{hod03:pasp115}) for our two northernmost targets, WD J1622+5840 and WD J2100+2122, as part of program GN-2019A-Q-303. The data were processed using DRAGONS (Data Reduction for Astronomy from Gemini Observatory North and South) v2.1.0, and the photometry was calibrated using nearby sources from the 2MASS All-Sky Point-Source Catalog (2MASS; \citealt{skr06:aj131}) as reference stars. The NIRI \emph{J}, \emph{H}, and \emph{K} filters are based on the Mauna-Kea Observatories (MKO) near-infrared system \citep{sim02:pasp114}, and published color transformations were used to transform the 2MASS photometry to the MKO system prior to determining a zero-point for each combined image \citep{leg06:mnras373, hod09:mnras394}. 

\emph{Gemini-S/Flamingos-2}:
\emph{J}, \emph{H}, and \emph{Ks}-band photometry was collected using the Flamingos-2 near-infrared wide-field imager and spectrograph at Gemini-South (Flamingos2; \citealt{eik04:spie5492}) for the three remaining targets as part of programs GS-2018B-FT-204 and GS-2018B-Q-404. The data were processed using DRAGONS v2.1.0, and and the photometry was calibrated using nearby 2MASS sources as reference stars. The Flamingos-2 \emph{J} and \emph{H} filters are based on the MKO filter set, while the \emph{Ks} filter is similar to the 2MASS \emph{Ks} filter \citep{leg15:apj799}, and the same color transformation were used to transform the 2MASS \emph{J} and \emph{H} photometry for the calibration sources prior to determining a zero-point for each combined image. 

\emph{Spitzer/IRAC \label{sec:spit}}:
We collected 3.6 and 4.5 $\mu$m photometry using the Infrared Array Camera on \emph{Spitzer} as part of program 14220 (IRAC; \citealt{faz04:apjs154}). Eleven frames were taken using 30\,s exposures with the medium-sized cycling dither pattern, resulting in 330\,s of total integration in each channel. We produced fully calibrated mosaic images for each target using the MOPEX software package \citep{mak06:spie6274} following the recipes outlined for point-source extraction in the \emph{Spitzer} Data Analysis Cookbook v6.0. PRF-fitted photometry was conducted on each target using APEX, with the appropriate corrections applied. 

WD J0611$-$6931 was not included as part of this program, but fell within the field-of-view of program 70062 of Cycle 10 (PI: J. Davy Kirkpatrick). The target is identified near the edge of the dither pattern in IRAC Ch$\,$1, presenting some doubt of the measurement as there are several nearby visual artifacts in the fully reduced CBCD frames. We performed aperture photometry on the CBCD frames with a small aperture to avoid the visual artifacts. Aperture corrections were applied as recommended in the IRAC Instrument Handbook.

\section{White Dwarf Atmospheric Properties and Signs of Accretion \label{sec:wd}}

\begin{figure}
\gridline{\fig{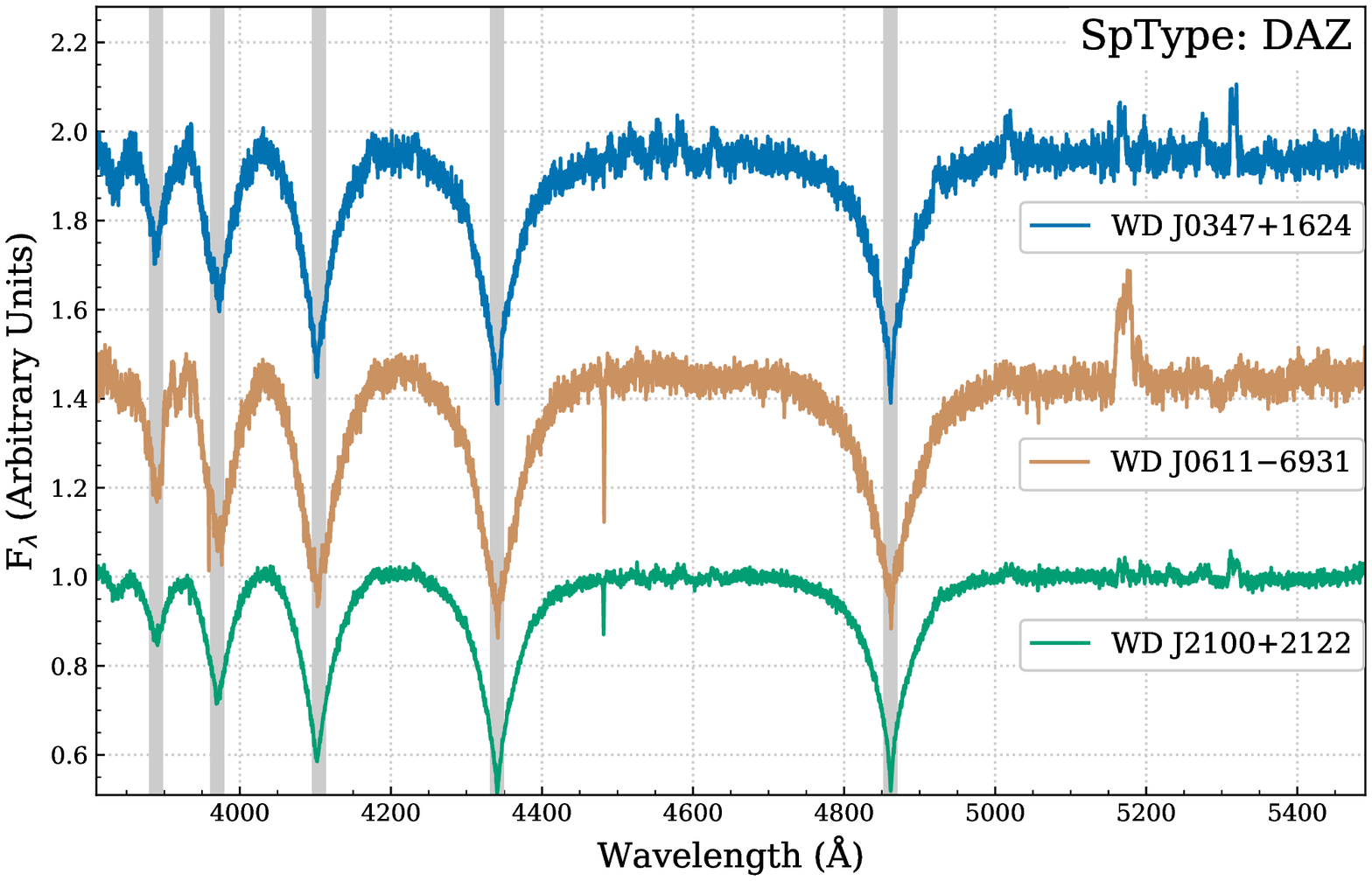}{0.45\textwidth}{}}
\vspace{-0.8cm}
\gridline{\fig{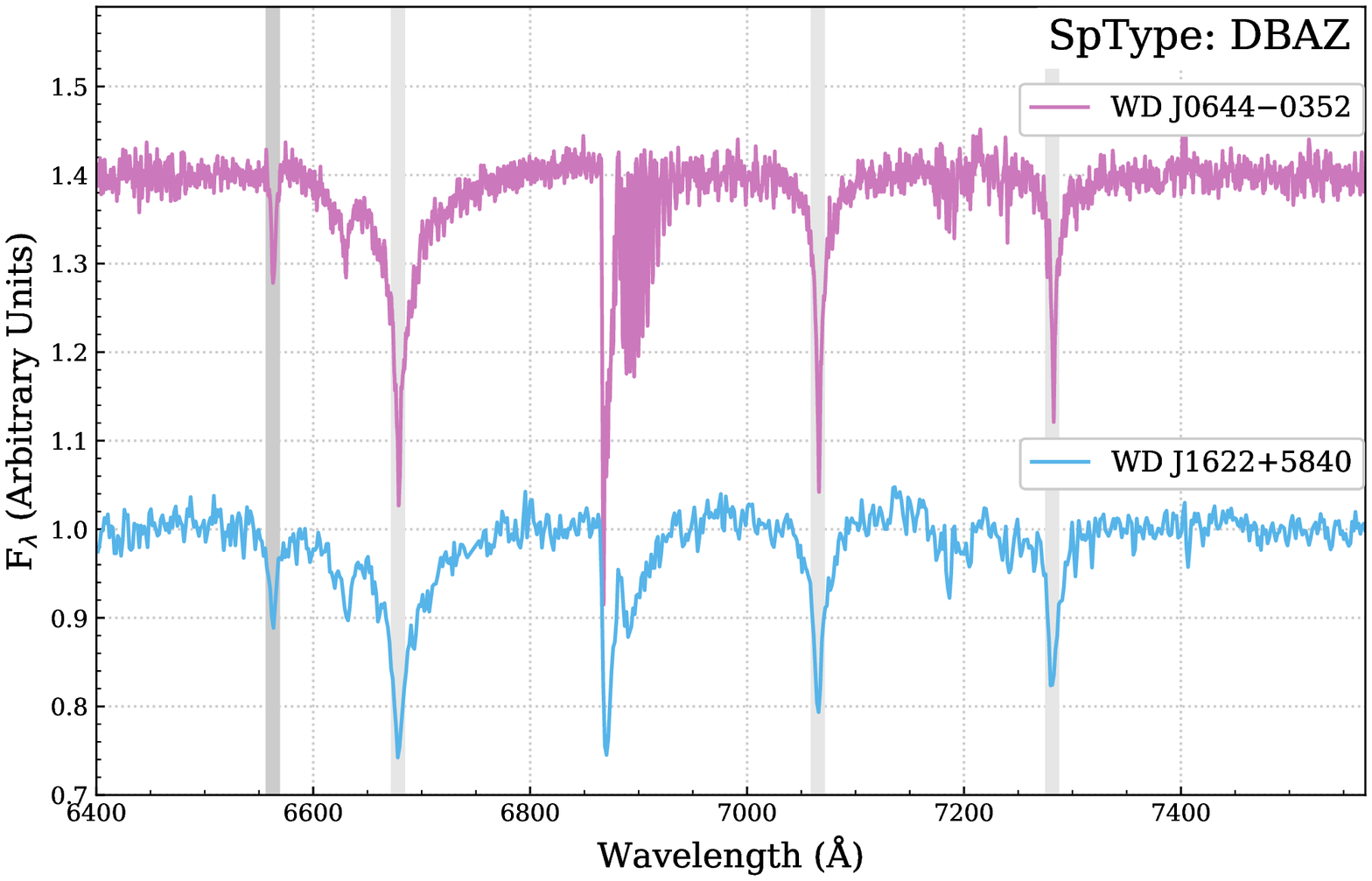}{0.45\textwidth}{}}
\caption{Normalized spectra for all five targets, separated by spectral type. The top four spectra were taken with VLT/X-Shooter, while the bottom spectra is from Gemini-N/GMOS. The three hydrogen-dominated atmosphere white dwarf stars are identifiable in the top panel by their broad balmer series absorption features (dark grey bands), and additional emission species from \ion{Fe}{2} and \ion{Mg}{1} are also seen in all three DAZ spectra, as well as \ion{Mg}{2} absorption near 4481\,\AA in the bottom two spectra. The two helium-dominated atmosphere white dwarf spectra are shown in the bottom panel, with \ion{He}{1} features identified in light grey. Telluric features are present in the two DBAZ spectra between 6850 and 6950\,\AA.\label{fig:fullspec}}
\end{figure}

Our broad-band and high-resolution spectroscopic follow-up for each target serves to characterize the white dwarf stellar atmosphere and to search for signs of ongoing accretion from the circumstellar material. For stellar classification, we present our broad-band spectra in Figure \ref{fig:fullspec} and the spectral type and atmospheric parameters of each target are given in Table \ref{tab:wdspec}. The five systems span a broad range of stellar parameters, but we note the particularly high effective temperature of the white dwarf star in WD J2100+2122. At just over 25,000$\,$K, it is the hottest white dwarf known to host dust and gaseous debris in emission (see Table 2 of \citealt{man16:mnras462}), and the effects of this high stellar temperature on the emission spectra are discussed below. 

\begin{figure}
\gridline{\fig{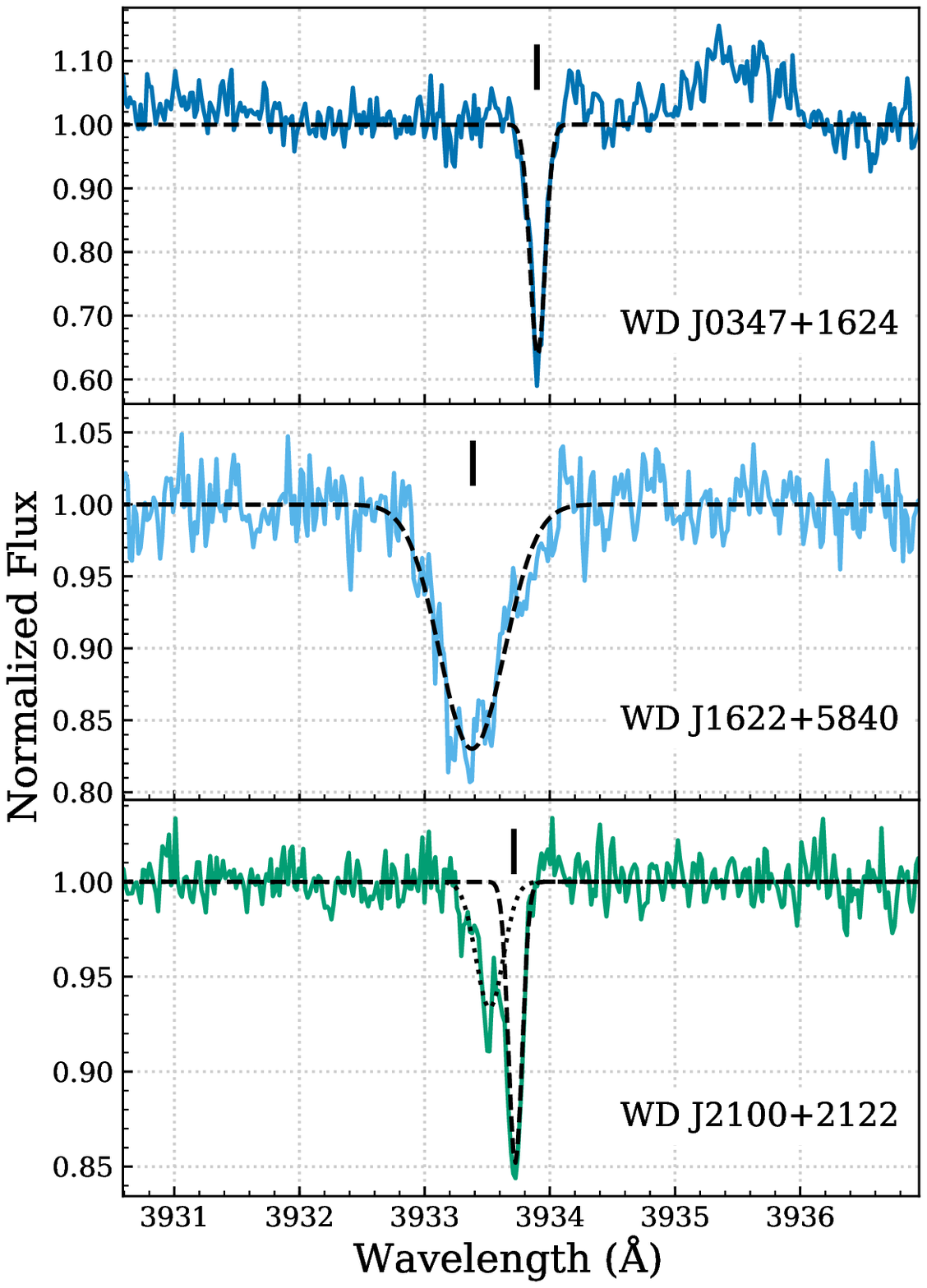}{0.4\textwidth}{}}
\vspace{-0.8cm}
\gridline{\fig{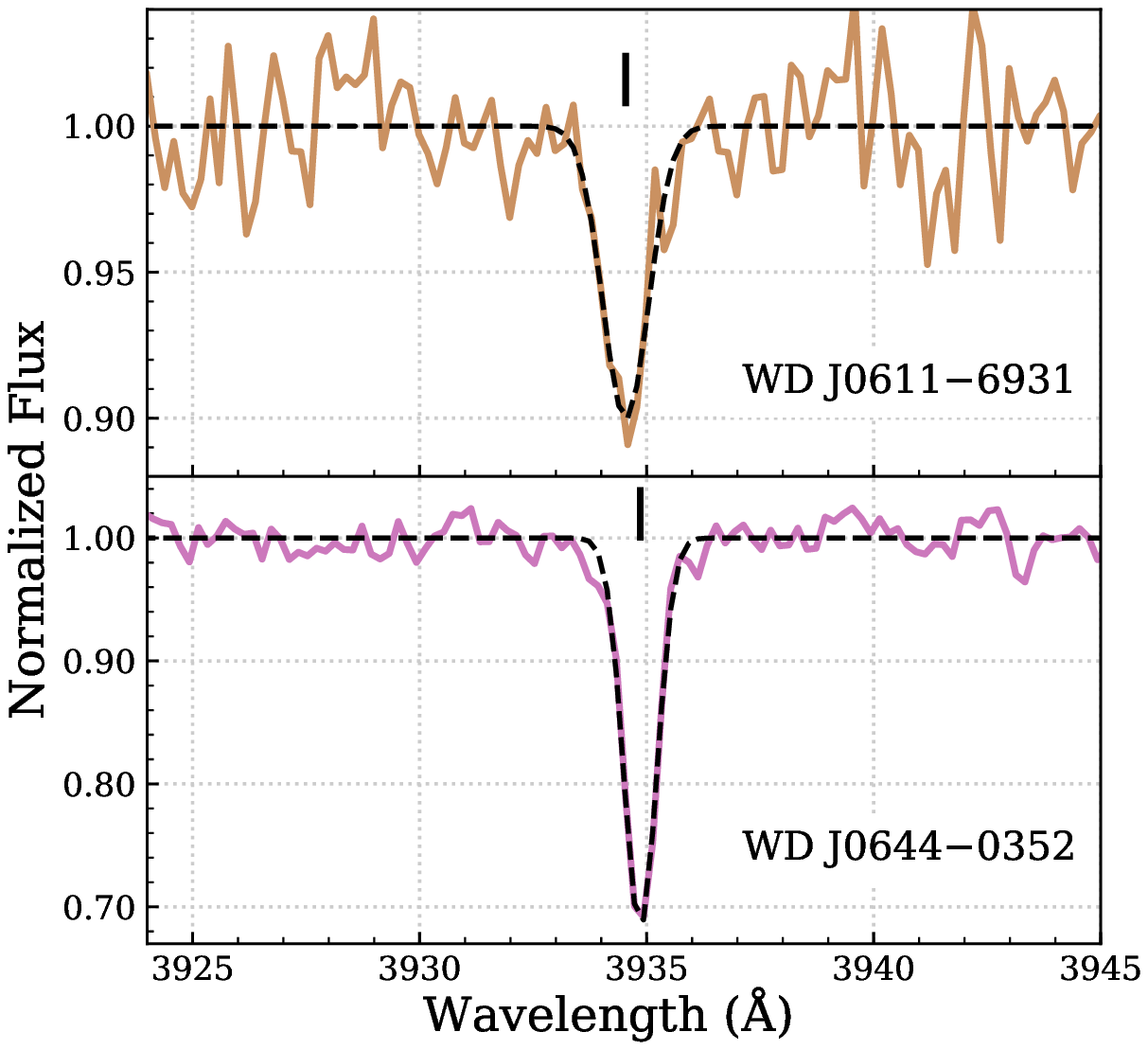}{0.4\textwidth}{}}
\caption{Atmospheric Ca K absorption is observed in all five targets using echelle spectra collected from Keck/HIRES (top panel) and VLT/X-Shooter (bottom panel). WD J2100+2122 shows evidence of multiple absorption components while the absorption profile for WD J1622+5840 is significantly broadened. The dashed black line shows our Gaussian profile fits and the vertical black dash denotes the measured centroid of the Ca K feature that is consistent with the white dwarf apparent velocity. The second Gaussian plotted as a dotted line in the WD J2100+2122 panel is likely due to interstellar abosorption. \label{fig:cak}}
\end{figure}

Ongoing atmospheric accretion is a key property of all known white dwarf stars that host debris disks \citep{kil06:apj646}. The most commonly detected feature at optical wavelengths for metal-polluted white dwarf stars is the Ca K feature at 3934\,\AA\ \citep{zuc03:apj596}. We produce normalized cutouts of the region surrounding this feature (Figure \ref{fig:cak}), and detect narrow Ca K absorption features in the spectra of all five systems. While additional metal species are detected in some targets (e.g. Mg, Fe, Si), we leave the detailed atmospheric abundance studies for future works. 

To confirm the Ca K is atmospheric, we compare its radial velocity ($v_{\rm{Ca}}$) to the apparent white dwarf stellar velocity ($v_{\rm{app}}$) in the heliocentric corrected reference frame. The apparent velocity of the the white dwarf star includes contributions from a gravitational redshift due to the strong surface gravity and the systemic velocity. For the stars with hydrogen-dominated atmospheres (WD J0347+1624, WD J0611$-$6931 and WD J2100+2122), we measure the apparent velocity by fitting a Gaussian profile to the sharp, non-local thermodynamic equilibrium (NLTE) cores of the hydrogen Balmer series. At the signal-to-noise of our observations, both the H$\alpha$ and H$\beta$ NLTE line cores are detected. For the stars with helium-dominated atmospheres (WD J0644$-$0352 and WD J1622+5840), the \ion{He}{1} features at 4471\,\AA\ and 5876\,\AA\ are likewise fit with a Gaussian profile and all available measurements are combined to produce the final $v_{\rm{app}}$.

The individual line measurements and the epoch of the spectra used to measure the radial velocities are given in Table \ref{tab:wdspec}. In white dwarf atmospheres, differential pressure shifts can result in radial velocity differences on the order of a few km\,s$^{-1}$; see \citet{fal12:apj757} for a discussion on this effect on measurements of the apparent velocities of white dwarf stars with both hydrogen and helium features visible. We do not account for these systematic differences in the reported velocities in Table \ref{tab:wdspec}, but note that they are on the order of our uncertainties.

\begin{deluxetable*}{lllcccccccc}
\tablecaption{Atmospheric properties of white dwarf stars in our sample \label{tab:wdspec}}
\tablehead{\colhead{Name} & \colhead{SpType} & \colhead{T$_{\rm eff}$} &
\colhead{$\log {g}$} & UT Date & \colhead{$v_{\rm{H}\alpha}$} & \colhead{$v_{\rm{H}\beta}$} & \colhead{$v_{\rm{He}I}$} &  \colhead{$v_{\rm{He}I}$} & \colhead{$v_{\rm{app}}$} & \colhead {$v_{\rm{Ca}}$} \\ \colhead{} & \colhead{} & \colhead{(K)} &
\colhead{(cm\,s$^{-2}$)} & \colhead{} & \colhead{6562.7} & \colhead{4861.3} & \colhead{5875.6} &  \colhead{4471.5} & \colhead{} & \colhead{3933.7}}
\startdata
WD J0347+1624  & DAZ &  20,620  &  7.89 & 2019 Dec 05 & - &  27$\pm$8  & - & - & 27$\pm8$ & 18$\pm1$\\ 
WD J0611$-$6931  &  DAZ & 16,550  &  7.87 & 2019 Oct 15 & 70$\pm$2 & 62$\pm$5 & - & - & 68$\pm5$ & 67$\pm5$\\ 
WD J0644$-$0352  &  DBAZ & 20,850  &  8.17 & 2019 Sep 14 & 99$\pm$4 & - & 95$\pm$4 &  104$\pm$5 &  98$\pm$7 & 92$\pm$2\\ 
WD J1622+5840  & DBAZ & 19,560  &  7.88 & 2019 Jul 12 & - & - & -17$\pm$2 & -18$\pm$3 & -17$\pm$4 & -21$\pm$1\\ 
WD J2100+2122  &  DAZ & 25,320  &  8.07 &  2019 Jul 10 & - & 6$\pm$2 & - & - &  6$\pm$2 & 5$\pm$1\\ 
\enddata
\tablecomments{White dwarf stellar effective temperature T$_{\rm eff}$ and surface gravity $\log {g}$ are taken from \citet{gen19:mnras482}. All velocities are given in units of km\,s$^{-1}$. For each velocity measurement we give the reference wavelength in air of the lines used in units of \AA, taken from the NIST Atomic Spectral Database.}
\end{deluxetable*}

To measure the radial velocity of the Ca K absorption feature, we fit a Gaussian profile to the normalized spectra shown in Figure \ref{fig:cak}. The five systems in our sample exhibit a wide range of morphology in their Ca K absorption profiles. In fact, WD J0644$-$0352 is the only target that appears to have a simple absorption profile that is well-described by a single Gaussian function. The Ca K absorption features in both WD J0347+1624 and WD J0611$-$6931 are contaminated by broad, double-peaked emission from the gaseous debris, affecting our continuum normalization and Gaussian fits. This systematic uncertainty is not captured in our reported measurements and the measured Ca K velocity for both systems agrees with the measured white dwarf apparent velocity within the reported uncertainties. 

The Ca K absorption feature observed in the spectra of WD J2100+2122 has a clear two-component structure. To determine the  separation of the two components, we simultaneously fit two Gaussian profiles to the data, shown in the third panel of Figure \ref{fig:cak} as a dashed and dotted line. The central wavelengths of the two features are 3933.51, and 3933.72\,\AA, corresponding to radial velocities of $-$11 and +5 km\,s$^{-1}$. The apparent velocity of the white dwarf star is +4.9 km\,s$^{-1}$, suggesting only the primary absorption component (dashed line) is atmospheric. We use the Local Interstellar Medium (ISM) model of \citet{red08:apj673}\footnote{\url{http://lism.wesleyan.edu/LISMdynamics.html}} to search of known ISM clouds along the line-of-sight for WD J2100+2122. The line-of-sight intercepts two known ISM clouds, Eri $v_r=-$11.79 km\,s$^{-1}$ and Aql $v_r=-$4.74 km\,s$^{-1}$. This indicates that the $-$11 km\,s$^{-1}$ component (dotted line) is likely due to ISM absorption from the Eri cloud. 

WD J1622+5840 has a wide absorption feature, though individual components are not resolved as in WD J2100+2122. A single Gaussian fit to the Ca K profile results in a radial velocity that is consistent with the apparent white dwarf velocity, indicating the feature is primarily atmospheric. Yet, with a full-width half-maximum (FWHM) of 0.61\,\AA, the absorption profile is much broader than the atmospheric  features we detect in WD J0347+1624 and WD J2100+2122 (0.14\,\AA\, and 0.13\,\AA\, respectively), both of which are consistent with instrumental broadening at this wavelength (0.11\,\AA\, at 3934.66\AA). It is clear that there is something additional contributing to the width of this absorption profile.

A search of the \citet{red08:apj673} database reveals no known ISM clouds along the line-of-sight. Furthermore, metal lines of Ca H, \ion{Si}{2}, and \ion{Mg}{2} are detected and show evidence of similar broadening, which would we do not expect from interstellar gas. Observations of nearby sources along the same line-of-sight could definitively rule out contributions from the ISM. Circumstellar gas, such as what was detected in the transiting material around WD 1145+017\citep{xu16:apjl816}, is another option. However the absorption morphology of the circumstellar lines seen in WD 1145+017 is very different than what we see in WD J1622+5840, with lines that are much shallower and often asymmetric. Time evolution of the absorption morphology would strongly indicate circumstellar material, as has been seen for WD 1145+017 \citep{cau18:apjl852, for20:apj888}.   

Another intriguing interpretation is broadening due to Zeeman-splitting from a magnetic field at the white dwarf surface. Given the proximity of the circumstellar debris to the white dwarf star, weak magnetic fields can have a strong effect on circumstellar material as the alfv\'en radius for the magnetic capture of in-falling material strongly overlaps with the debris disk region (see Figure 5 of \citealt{far18:mnras474}). For a rough estimate of the magnetic field strength ($B_{*}$), we compared the measured FWHM of the Ca K absorption feature to simulated spectra of varying magnetic field strengths assuming the stellar parameters listed in Table \ref{tab:wdspec} and a fixed calcium abundance. At these line strengths, increasing the calcium abundance can increase the line depth, but has little effect on the line width. We find that the width of the Ca K feature is most consistent with a magnetic field strength between $20$\, and $30$\,kG. We do not detect Zeeman-splitting in any of the hydrogen or helium lines observed, however our resolution is not sufficient to rule out splitting of such a weak field. Higher resolution spectra covering these features or spectropolarimetric measurements could be used to confirm the weak magnetic field. If confirmed, the magnetic field of the white dwarf star is likely to have a strong influence on the inner edge of the debris disk. 

\section{Spectral Energy Distributions and Dusty Infrared Excesses \label{sec:ir}}

\begin{figure*}[ht!]
\gridline{\fig{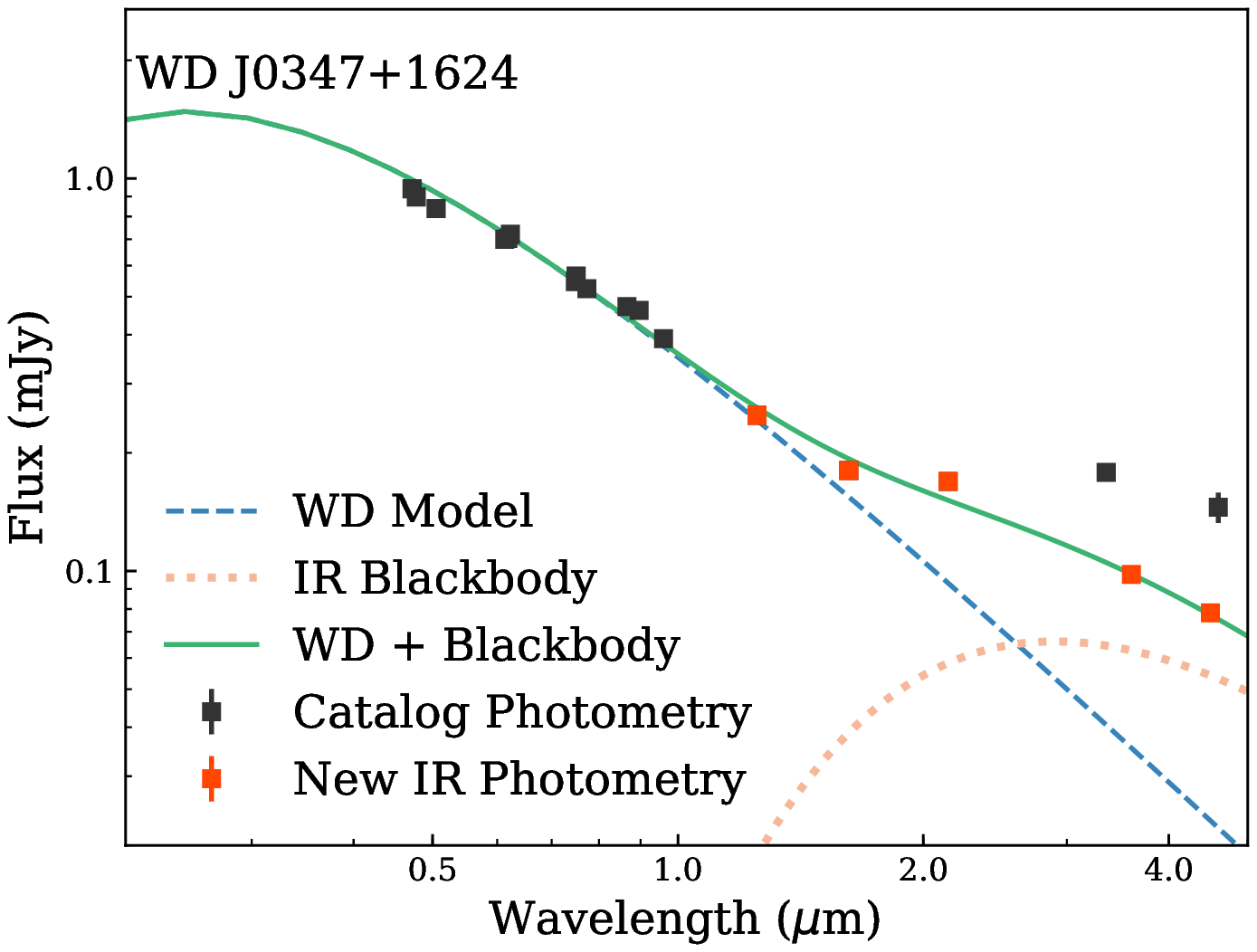}{0.34\textwidth}{ }\fig{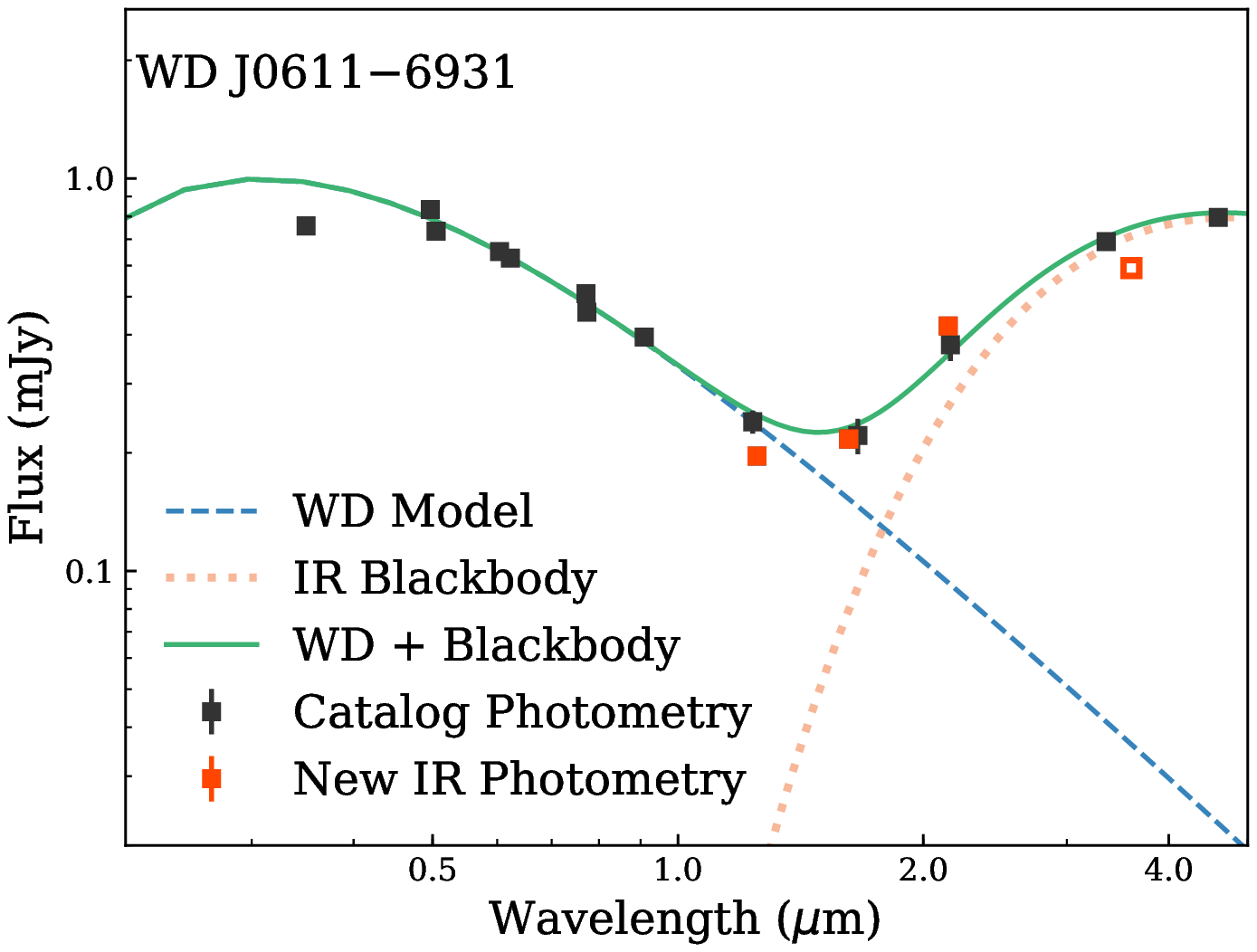}{0.34\textwidth}{}\fig{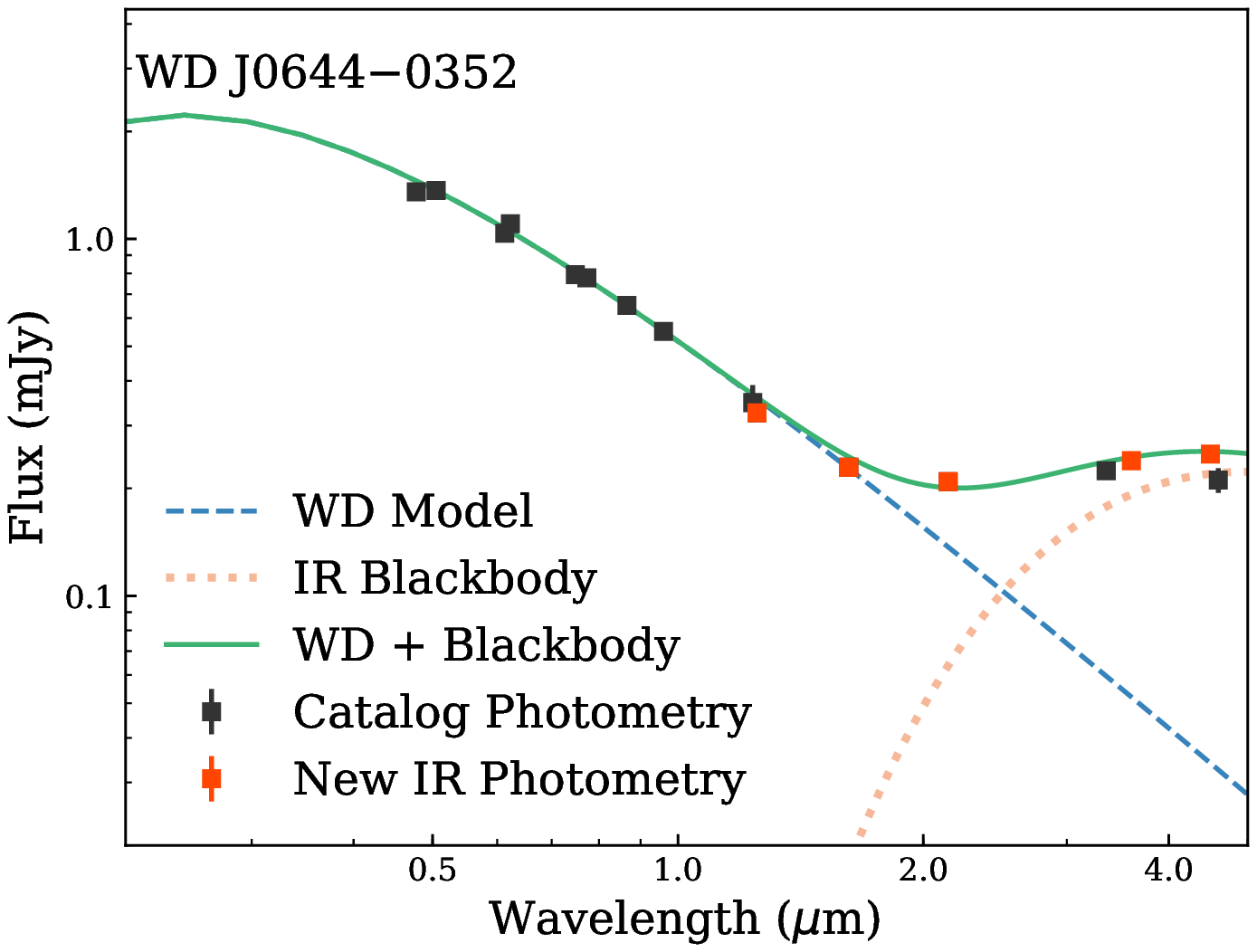}{0.34\textwidth}{}}
\vspace{-0.8cm}
\gridline{\fig{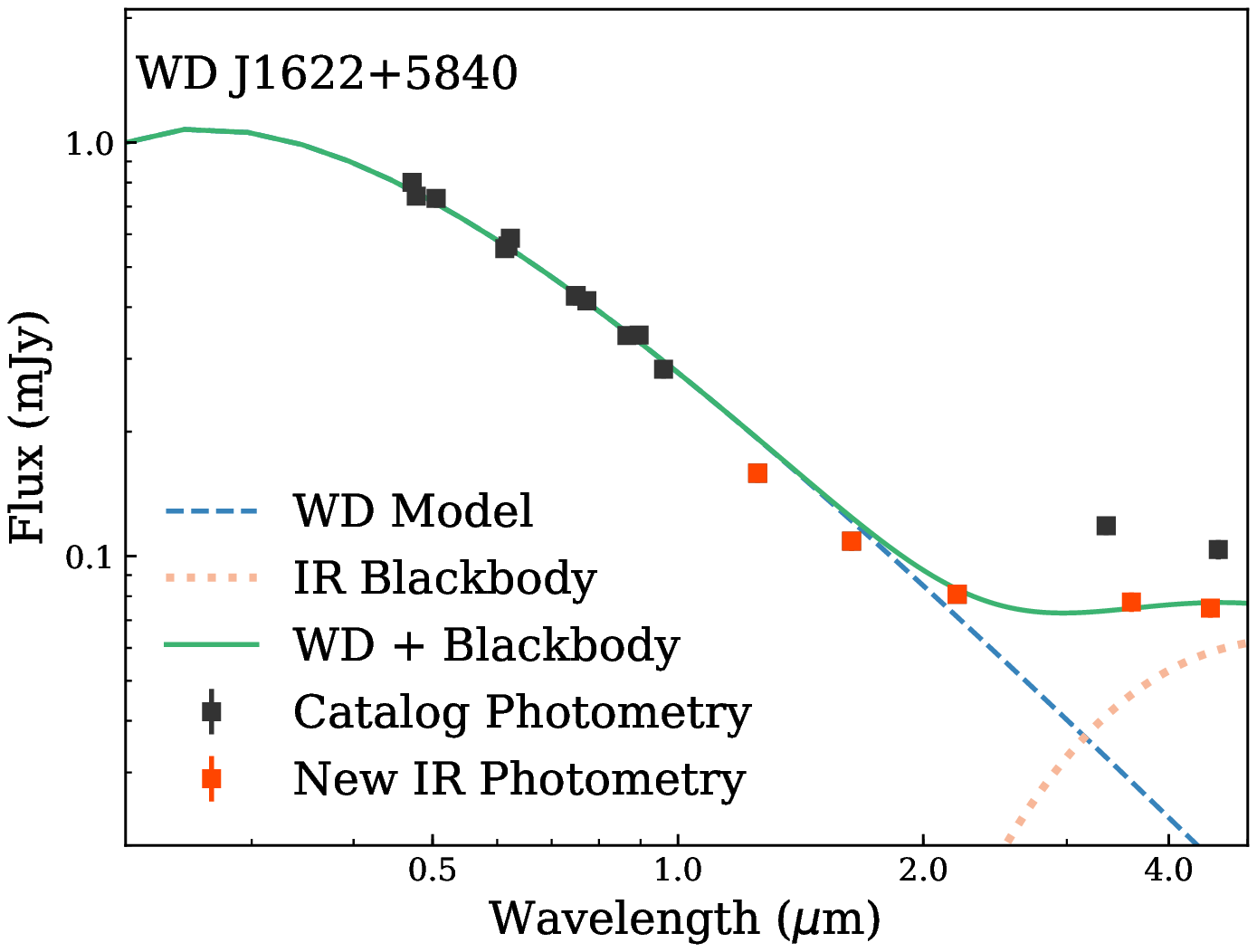}{0.34\textwidth}{}\fig{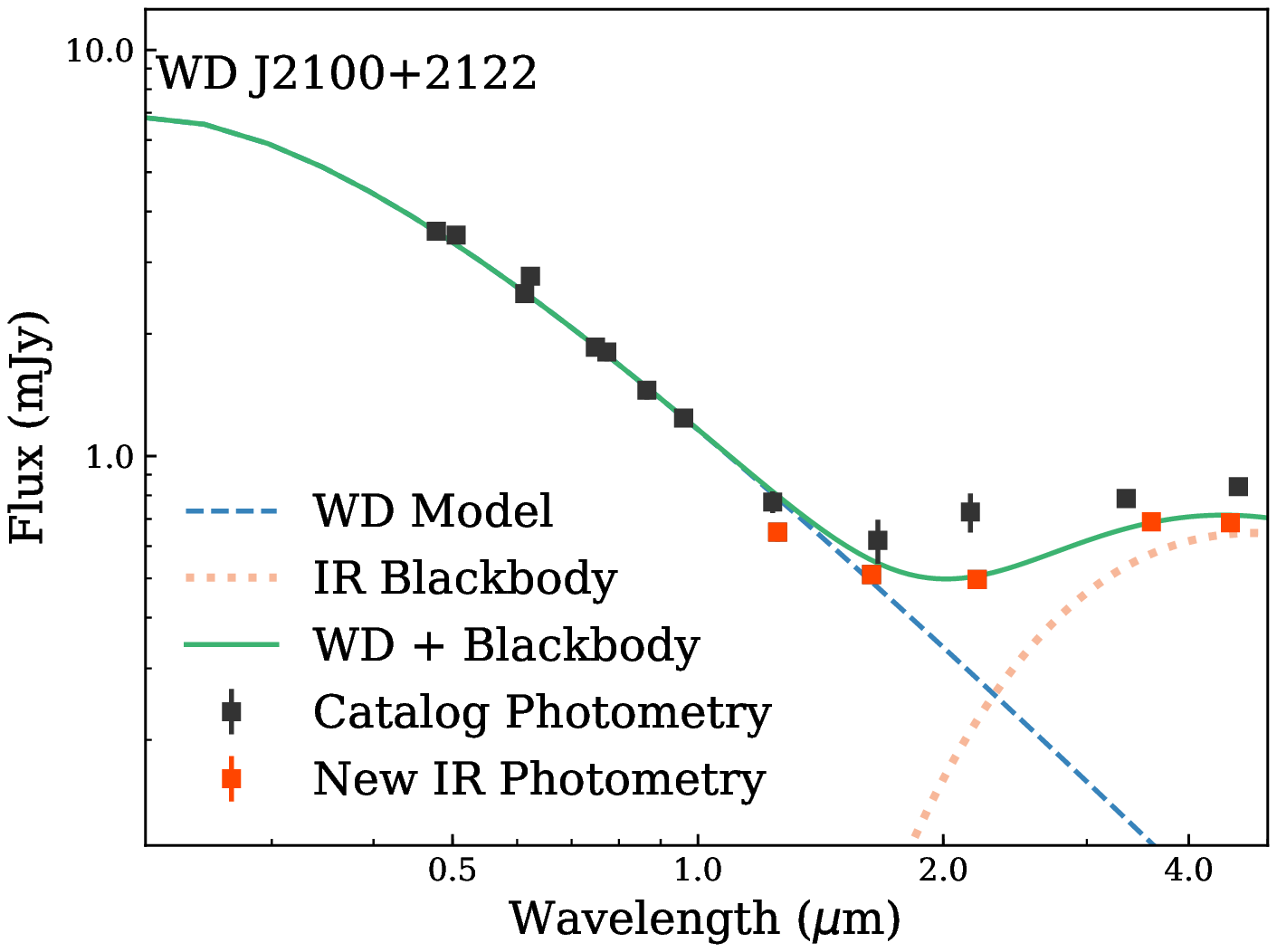}{0.34\textwidth}{}}
\caption{Spectral energy distributions of all five targets show strong infrared excesses attributable to compact, dusty debris disks. The white dwarf stellar contribution (dashed line) is represented by a blackbody source assuming the stellar parameters listed in Table \ref{tab:wdspec}, and a second blackbody source (dotted line) is added to white dwarf flux to describe the observed excess infrared radiation (solid line). A few targets show discrepancies between the \emph{WISE} catalog photometry (black squares) and \emph{Spitzer} follow-up photometry (red squares), potentially indicating contamination in the \emph{WISE} bands or intrinsic variability. For fitting the infrared excess, we only consider the targeted follow-up photometry from \emph{Spitzer} and Gemini (red squares), with the exception of WD J0611$-$6931 which may have a spurious \emph{Spitzer} Ch$\,$1 measurement (open square). \label{fig:sed}}
\end{figure*}

\begin{deluxetable*}{lcccccccccccc}
\tablecaption{Infrared properties of the dusty debris detected in our sample. \label{tab:wdphot}}
\tablehead{\colhead{Name} & Gaia \emph{G} & \multicolumn3c{Gemini} & \multicolumn2c{\emph{Spitzer}} & \multicolumn2c{All\emph{WISE}} & \colhead{T$_{\rm IR}$} & \colhead{R$_{\rm IR}$} & \colhead{$\tau$} \\ \colhead{} & \colhead{(mag)} & \colhead{\emph{J}} & \colhead{\emph{H}} & \colhead{\emph{K}} & \colhead{Ch$\,$1} & \colhead{Ch$\,$2} & \colhead{\emph{W1}} & \colhead{\emph{W2}} & \colhead{(K)} & \colhead{(R$_{\rm WD}$)} & \colhead{(\%)}}
\startdata
WD J0347+1624 &  16.7  & 249$\,\pm\,$13  &  180$\,\pm\,$10 &  169$\,\pm\,$9 & 98$\,\pm\,$5 & 78$\,\pm\,$4 & 178$\,\pm\,$9 & 145 $\,\pm\,$13 & 1760 & 9 & 0.38\\ 
WD J0611$-$6931 &  16.8 &  196$\,\pm\,$10 & 216$\,\pm\,$10  &  421$\,\pm\,$21 & 591$\,\pm\,$30 & -  & 690$\,\pm\,$35 & 796$\,\pm\,$40 & 1070 & 55 & 5.16\\ 
WD J0644$-$0352  & 16.2 & 325$\,\pm\,$16  & 229$\,\pm\,$11  &  208$\,\pm\,$10 & 239$\,\pm\,$12 & 249$\,\pm\,$13 & 224$\,\pm\,$11 & 210$\,\pm\,$17 & 1030 & 29 & 0.49 \\ 
WD J1622+5840 &  16.9 & 159$\,\pm\,$10  & 111$\,\pm\,$7  &  83$\,\pm\,$4 & 78$\,\pm\,$4 & 75$\,\pm\,$4 & 118$\,\pm\,$6 & 104$\,\pm\,$6 & 900 & 25 & 0.27 \\ 
WD J2100+2122  & 15.2 & 662$\,\pm\,$33  & 529$\,\pm\,$27  &  507$\,\pm\,$25 & 689$\,\pm\,$35 & 685$\,\pm\,$34 & 786$\,\pm\,$39 & 840$\,\pm\,$42 & 1060 & 36 & 0.39 \\
\enddata
\tablecomments{All infrared fluxes are given in units of micro-Janskys. Note that the WD J0611$-$6931 \emph{Spitzer} Ch$\,$1 measurement may be spurious (see Section \ref{sec:spit}).}
\end{deluxetable*}

The dusty components of debris disks result in spectral energy distributions with a characteristic excess of infrared radiation over the white dwarf stellar flux \citep{jur03:apjl584,kil06:apj646, von07:apj662}, although some circumstellar debris disks around white dwarf stars are known to escape detection in the infrared \citep{ber14:mnras444,koe14:aap566,bon17:mnras468}. 

To construct the spectral energy distributions for our targets and search for this infrared excess, we collected publicly available photometry from the Sloan Digital Sky Survey \citep{ahn14:apjs211}, Panoramic Survey Telescope and Rapid Response System \citep{cha16:arxiv}, SkyMapper Southern Survey \citep{wol18:pasa35}, Two-Micron All-Sky Survey (2MASS; \citealt{skr06:aj131}), and the All\emph{WISE} surveys \citep{cut13:ycat2328}. We de-reddened the photometry using the technique described in \citet{gen19:mnras482}, and converted the magnitudes into units of flux density using the published zero-points for each bandpass. In addition, we collected targeted infrared photometry as described in Section \ref{sec:photfollow}. The measured fluxes were converted in units of flux density and are presented in Table \ref{tab:wdphot}. The resulting spectral energy distributions combine our targeted follow-up with the collected survey data, and are shown in Figure \ref{fig:sed}. 

\subsection{Establishing The Infrared Excess}

To model the stellar contribution to the measured photometry, we approximate the white dwarf star with a blackbody source with a fixed effective tempearture and radius consistent with the parameters given in Table \ref{tab:wdspec}. The radius is derived from the $\log {g}$ value using the evolutionary tracks of \citealt{fon01:pasp113} with the appropriate spectral type. The stellar model is then scaled to the observed flux using all photometry measurements with wavelengths below one micron. Approximating the white dwarf star as a blackbody source is not appropriate for detailed modeling of the infrared excess, but serves our purpose as a qualitative point of reference for the stellar contribution to the infrared flux. All five targets exhibit significant excess infrared radiation above the expected stellar contribution beyond 1.5 microns.
 
Infrared excess candidates that depend on \emph{WISE} photometry are well known to suffer from appreciable amounts of contamination \citep{bar14:apj786,sil18:apj868}, that in many cases require \emph{Spitzer} follow-up to resolve \citep{den20:apj891}. In this case, all five targets were initially selected for follow-up based on their \emph{WISE} excess (\citealt{reb19:mnras489}; Xu et al. 2020, submitted to ApJ), necessitating a careful comparison of the catalog photometry and the infrared follow-up photometry. Three targets in our sample, WD J0347+1624, WD J1622+5840 and WD J2100+2122, showed significant discrepancies between the catalog and collected infrared photometry as seen in Figure \ref{fig:sed}. In all three systems the collected \emph{Spitzer} photometry is lower than the reported \emph{WISE} photometry, which could be reconciled by additional flux in the \emph{WISE} measurements due to source confusion \citep{wil17:mnras468,den20:apj891}. 

To help distinguish between source confusion and intrinsic variability, we examine the \emph{Spitzer} Ch$\,$1 and Ch$\,$2 frames for evidence of nearby sources. We find that WD J1622+5840 is blended with a nearby source in both Ch$\,$1 and Ch$\,$2 images that is only resolved through PRF-fitted photometry. We perform aperture photometry on the \emph{Spitzer} images with a radius of 7.8\arcsec, corresponding to the approximate confusion limit of the automatic source de-blending routine used in All\emph{WISE} pipeline\footnote{\url{http://wise2.ipac.caltech.edu/docs/release/allsky/expsup/sec4_4c.html}}. The wide aperture photometry of the \emph{Spitzer} Ch$\,$1 and Ch$\,$2 is consistent with the photometry reported by \emph{WISE}, confirming source confusion is responsible for the difference in flux seen in WD J1622+5840. 

The \emph{Spitzer} images for WD J0347+1624 and WD J2100+2122 show the targets are clear of nearby contaminating sources, and large aperture photometry of the IRAC Ch$\,$1 and Ch$\,$2 images does not resolve the discrepancy between the All\emph{WISE} and \emph{Spitzer} photometry. Recently, infrared variability of dusty debris disks around white dwarfs has been detected on timescales of months and years \citep{xu18:apj866, far18:mnras481, swa19:mnras484, wang19:apjl886}. Intrinsic variability remains a plausible explanation for these two targets, and presents some issues for our attempts to fit or describe the infrared excess, as the spectral energy distributions presented in Figure \ref{fig:sed} include survey data with epochs spanning several years. To attempt to mitigate the effects of intrinsic variability on our fits to the infrared excess, we choose to only consider the newly collected infrared photometry, which was all collected within the past two years. 

As discussed in Section \ref{sec:photfollow}, the \emph{Spitzer} data for WD J0611$-$6931 is potentially spurious, so we choose to rely on the \emph{WISE} photometry when fitting the infrared excess. In the IRAC Ch$\,$1 images, the target is free of nearby sources that might contaminate the \emph{WISE} photometry, and there is a rough agreement between the IRAC Ch$\,$1 and \emph{WISE} \emph{W1} fluxes, supporting this choice.   

\subsection{Characterizing The Infrared Excess}

To describe the infrared excess, we fit an additional single-temperature blackbody component to the spectral energy distribution using a chi-squared minimization process with the temperature and radius set as free parameters and the blackbody source placed at the distance of the white dwarf star. We show the best-fitting combination of white dwarf and blackbody source along with the collected photometry in Figure \ref{fig:sed}, and report the parameters of the best-fitting infrared source in Table \ref{tab:wdphot}.

\begin{deluxetable*}{lcccccc}[ht!]
\tablecaption{Properties of the emission features from the gaseous debris detected in our sample. \label{tab:emission}}
\tablehead{\colhead{Name} & \colhead{UT Date} & \colhead{Ca 8498 Eqw} & \colhead{Ca 8542 Eqw} & \colhead{Ca 8662 Eqw} &
\colhead{FWZI} & \colhead{Additional Emission Species} \\ \colhead{} & \colhead{} & \colhead{(\AA)} & \colhead{(\AA)} & \colhead{(\AA)} &\colhead{(km\,s$^{-1}$)} & \colhead{}}
\startdata
WD J0347+1624  & 2018 Dec 03 & 5.24$\pm$0.4  &  5.61$\pm$0.39  &  6.52$\pm$0.4 & 630$\pm$50 & \ion{O}{1}, \ion{Fe}{2} \\ 
 			  & 2019 Jan 24 & 5.9$\pm$0.13  &  6.57$\pm$0.13  &  6.31$\pm$0.13 & 670$\pm$50 & \\ 
			  & 2019 Nov 14 & 7.74$\pm$0.11  &  8.69$\pm$0.11  &  7.96$\pm$0.11 & 590$\pm$30 & \\ 
WD J0611$-$6931  & 2019 Oct 15 & 10.65$\pm$0.16  &  14.92$\pm$0.17  &  15.75$\pm$0.17 & 1360$\pm$50 & \ion{O}{1}, \ion{Si}{1}, \ion{Mg}{1}, \ion{Na}{1}, \ion{Fe}{2} \\ 
WD J0644$-$0352  & 2019 Feb 11 & 0.78$\pm$0.1  &  0.93$\pm$0.1  &  1.21$\pm$0.1 & - & None \\    
			  & 2019 Mar 23 & 0.85$\pm$0.1  &  0.91$\pm$0.1  &  1.06$\pm$0.1 & - &  \\ 
			  & 2020 Feb 14 & 1.02$\pm$0.14  &  1.14$\pm$0.13  &  1.23$\pm$0.13 & 910$\pm$70 &  \\ 
WD J1622+5840  & 2020 Mar 01 &  0.51$\pm$0.12  &  0.91$\pm$0.13  &  0.81$\pm$0.11 & - & \ion{O}{1} \\ 
 			   & 2020 May 03 &  0.72$\pm$0.10  &  1.03$\pm$0.12  &  1.02$\pm$0.12 & 1110$\pm$100 &  \\ 
WD J2100+2122  & 2019 Apr 09 & 0.26$\pm$0.1  &  0.25$\pm$0.1  &  0.34$\pm$0.1 & - & \ion{O}{1}, \ion{Fe}{2} \\ 
			  & 2019 May 15 & 0.53$\pm$0.27  &  0.43$\pm$0.28  &  0.81$\pm$0.27 & - &  \\ 
			  & 2019 Jun 17 & 1.83$\pm$0.42  &  2.83$\pm$0.43  &  1.61$\pm$0.42 & 650$\pm$70 &  \\ 
			  & 2019 Jul 12 & 0.44$\pm$0.12  &  0.27$\pm$0.12  &  0.37$\pm$0.12 & - &  \\ 
\enddata
\tablecomments{The FWZI measurements are an average of the three calcium infrared triplet emission profiles and are only calculated when a clear detection of both edges of the profile can be made. As discussed below, some systems show significant evolution over time in both profile strength and shape that lead to variations in these measurements.} 
\end{deluxetable*}

The temperature and radii of the best-fitting infrared source are not consistent with theoretical predictions for stellar and sub-stellar companions, as their radii are too large given their inferred temperatures \citep{cha97:aap327,cha00:apj542}. Given the detections of emission lines from circumstellar gas, the most likely interpretation is that the infrared excess is due to circumstellar dust. The inferred blackbody temperatures would correspond to dust re-radiating in thermal equilibrium that is near or within the tidal disruption radius of the white dwarf star, depending on whether the dust is assumed to be optically thick or optically thin \citep{jur03:apjl584,rea05:apjl635}. 

The optically thick and vertically thin model of \citet{jur03:apjl584} has since become the preferred description of the dust in these systems, and could help constrain the geometry of the dust. When combined with models of the gaseous emission lines, a more complete picture of the debris disk environment can be constructed, and the analyses of similar systems have shown the dusty and gaseous material to be largely coincident \citep{mel10:apj722}. However, we lack the longer wavelength observations needed to constrain models with multiple free parameters. The potential for intrinsic variability further complicates this effort, as recent works suggest several optically thin and thick components may be needed to describe such infrared variations \citep{swa20:mnras496}. We therefore choose to limit our characterization of the infrared excess to the best-fitting parameters of a single-temperature blackbody for this work. 

While a single-temperature blackbody is a poor physical representation of the circumstellar dust that is neither spherical nor at a single-temperature, it is still useful for comparison putting these discoveries in context of the known sample of white dwarf debris disks. Both \citet{roc15:mnras449} and \citet{den17:apj849} have explored the bulk sample properties of white dwarfs with known circumstellar debris disks modeled as single-temperature blackbodies. As compared with the sample analyzed in \citet{den17:apj849}, the best-fitting infrared temperatures and radii of the new debris disks are consistent with other white dwarf debris disks known to host gaseous debris in emission, strengthening the correlation between the brightness of the debris disk and its propensity to host gaseous debris in emission. 

One way to quantify the brightness of the debris disk is to calculate the fractional infrared brightness of the infrared source ($\tau$, given in units of percent), which we report in Table \ref{tab:wdphot}. By approximating the debris disk as a single-temperature blackbody, we are not capturing the true fractional infrared brightness, but it allows for direct comparison with the sample analyzed by \citet{roc15:mnras449}, who used similar methodology. The debris disks in all five white dwarfs show fractional infrared luminosities that are consistent with other dusty white dwarfs at similar temperatures \citep{roc15:mnras449}. WD J0611$-$6931 is an exception. The fractional infrared flux of 5.16 $\%$ is unusually high, exceeding that of GD 362, one of the brightest known dusty debris disks around a white dwarf \citep{bec05:apjl632,kil05:apjl632}. The strength of the infrared excess of GD 362 requires multiple dusty components to model \citep{jur07:aj133}, and a similar approach would be required if one considers all of the infrared flux from WD J0611$-$6931 to be a result of circumstellar dust. 

Finally, we emphasize that while instructive, our choice of a single-temperature blackbody models for fitting the spectral energy distribution limits the interpretations we can make. These simplifications allow us to quickly place the new discoveries in the context of other similar systems, but more detailed modeling is needed to understand the geometry and fundamental properties of the circumstellar dust.

\section{Gaseous Emission Line Shapes, Strengths, and Evolution \label{sec:gas}}

In addition to the excess infrared radiation from the dusty components of the debris disks around these stars, all five systems exhibit emission lines emanating from a gaseous component. Here we present the emission lines detected, including a search for variability over the first year of follow-up by combining data from Gemini/GMOS, VLT/X-shooter, and SOAR/Goodman.  Changes in the absorption and emission profiles of gaseous components in debris disks around white dwarf stars have been detected on timescales of minutes \citep{man19:sci364}, months \citep{red17:apj839}, years \citep{gan08:mnras391,cau18:apjl852, den18:apj854}, and decades \citep{wil14:mnras445,wil15:mnras451,man16:mnras455, man16:mnras462}, so the proper follow-up of newly discovered systems can be a daunting task. With only a few epochs for each system, we are unable to provide a detailed analysis, but our early follow-up hints at some familiar classes of variability and gives an opportunity to assess the future needs for each system. 

We focus our discussion on the calcium infrared triplet emission feature, as it is commonly the strongest emission feature detected in these systems \citep{man20:mnras493}. In Table \ref{tab:emission}, we provide a few measurables of the calcium triplet emission profiles for each system, including the equivalent width and the full-width zero-intensity (FWZI), measured as the separation between the points where the red/blue-shifted edges of the emission line profile meets the continuum. The reported FWZI value is the uncertainty weighted combination of the three independent FWZI measurements from the calcium triplet emission features. 

The FWZI and equivalent with measurements presented in Table \ref{tab:emission} are useful for comparison with known systems and for tracking the evolution of the emission features over long timescales. Our measurements are given at a single epoch but in our limited follow-up we already see signs of such evolution. In several of our systems we also note detections of Fe, Mg, Si, Na, and O in emission, presented in the Appendix. We discuss the emission species detected and the variability of each system individually below. 

\subsection{WD J0347+1624 \label{sec:0347}}

\begin{figure}[h!]
\epsscale{1.2}
\plotone{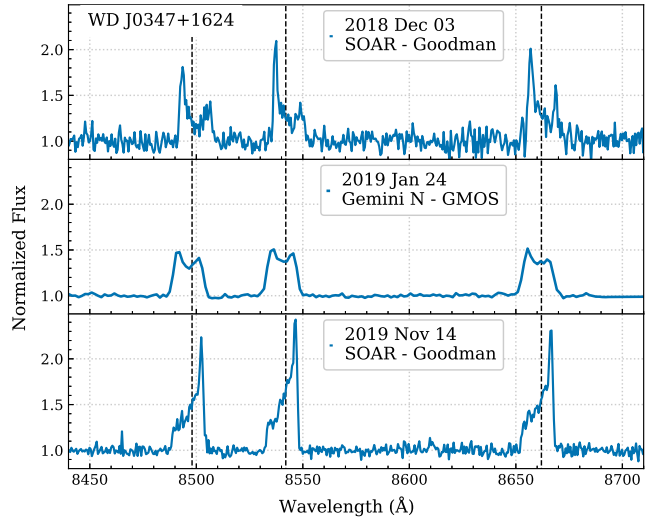}
\caption{The calcium infrared triplet emission line profiles in WD J0347+1624 are observed to undergo asymmetry shifts in our early follow-up, suggesting similar behavior to WD 1226+110 and HE 1349$-2305$. The dashed line marks the rest wavelength of the calcium infrared triplet. \label{fig:0347}}
\end{figure}

\begin{figure}[h!]
\epsscale{1.2}
\plotone{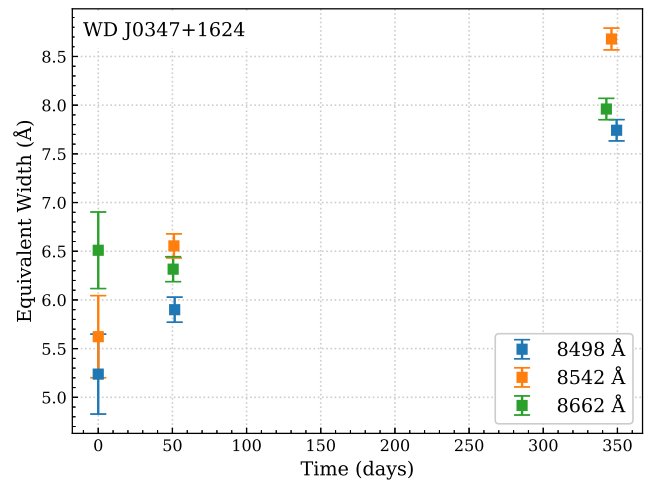}
\caption{The third epoch of our follow-up of WD J0347+1624 shows a significant increase in the equivalent widths of all three calcium triplet emission lines. \label{fig:wd0347eqw}}
\end{figure} 

The calcium infrared triplet emission from the gaseous debris around WD J0347+1624 was initially observed in December 2018 in a slightly asymmetric phase, with a blue-shifted peak as seen in Figure \ref{fig:0347}. The emission profiles are strong relative to other known white dwarf systems with gaseous debris in emission, reaching peak fluxes up to twice that of the white dwarf stellar continuum. With a full-width zero-intensity of just 630$\pm$40 km\,s$^{-1}$, the profiles are narrower than the similarly strengthened system WD 1226+110 \citep{gan06:sci314,mel10:apj722}, suggesting the disk is either observed at a lower inclination or that the emitting region is farther from the white dwarf star. 

Additionally, we see evidence of emission from \ion{O}{1} and \ion{Fe}{2} throughout the spectra, and emission at the Ca K transition 3934\,\AA, presented in Figure \ref{fig:wd0347_em}. Similar emission features have been observed in the spectra of the gaseous debris disks around Ton 345 \citep{mel10:apj722} and WD 1226+110 \citep{man16:mnras455}, and raise the possibility of performing direct compositional analyses of the gaseous material using spectral synthesis codes, such as what has recently been performed for the gaseous material around the white dwarf star WD J0914+1914 \citep{gan19:nat576}. Further discussion of the additional emission species detected amongst our sample is given in Section \ref{sec:discussion}. 

A follow-up spectrum in January 2019 showed a transition in the calcium infrared triplet emission to a more symmetric profile, and continued monitoring indicates the system is undergoing the same periodic global asymmetric profile evolution as has been recorded in the emission profiles of WD 1226+110 \citep{man16:mnras455} and HE 1349$-$2305 \citep{den18:apj854} and the absorption profiles of WD 1145+017 \citep{cau18:apjl852, for20:apj888}. If the emission profiles continue to evolve at the rate we have seen, the period for WD J0347+1624 is likely to be between two and three years. 

The periodic asymmetric evolution seen in these systems is well described by the precession of a fixed pattern within the disk \citep{har16:aap593, man16:mnras455}, driven by either general relativity or pressure differences within the disk \citep{mir18:apj857}. The recent detection of hourly variations in the emission profiles of WD 1226+110 suggests the presence of a planetesimal within the debris disk \citep{man19:sci364} that could be also responsible for the long-term emission profile behavior seen in WD 1226+110 and other similar systems \citep{man20:mnras493}. The strength and similarity in behavior of the calcium triplet emission line profiles of WD J0347+1624 make it a good test case for such theories. 

Finally, as shown in Figure \ref{fig:wd0347eqw}, the equivalent width of all three calcium triplet emission profiles in our 2019 Nov 11 spectra is significantly higher than the previous two epochs. The FWZI metric shows no significant variation over this time. It remains to be seen if this increase is stochastic or is correlated with the asymmetric profile evolution discussed earlier.

\subsection{WD J0611$-$6931 \label{sec:0611}}

\begin{figure}[h!]
\epsscale{1.2}
\plotone{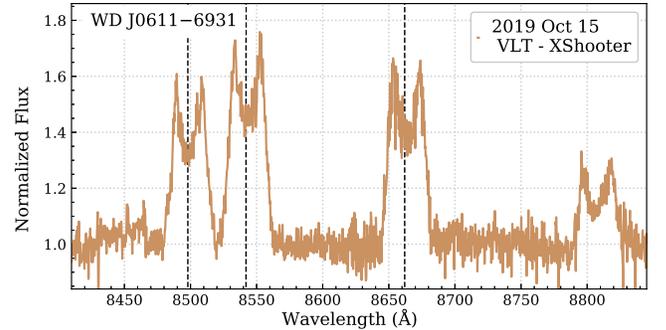}
\caption{Broad, mostly symmetric calcium triplet emission profiles are observed in WD 0611-6931, with an additional \ion{Mg}{1} emission feature seen at 8806\AA. \label{fig:0611}}
\end{figure}

The calcium infrared triplet emission from the gaseous debris around WD J0611$-$6931 was observed with VLT/X-Shooter in October 2019. As shown in Figure \ref{fig:0611}, the system displays strong and broad emission lines, similar in width to the gaseous debris of Ton 345 and SDSS1043+0855 \citep{gan07:mnras380, gan08:mnras391, mel10:apj722}. In addition to the calcium infrared triplet emission, we detect several other metal species in emission including lines from \ion{Mg}{1}, \ion{O}{1}, \ion{Fe}{2}, \ion{Si}{1}, and \ion{Na}{1}, shown in Figure \ref{fig:wd0611_em}. The latter two emission species are unique amongst the sample of white dwarfs with gaseous debris. All of the lines exhibit comparable width in velocity space, indicating the different species share a similar inner disk radius. 

In contrast to the additional metal species seen in the gaseous debris around the systems presented here and in the literature, the only \ion{Fe}{2} emission line we detect in the spectra WD J0611$-$6931 is at 5169\,\AA, likely blended with \ion{Mg}{1} emission. In addition, the \ion{Mg}{1} lines are comparable in strength to the calcium infrared triplet lines, which are typically much stronger than all other species \citep{man20:mnras493}. The strength of the \ion{Mg}{1} lines, the detection of \ion{Si}{1}, the lack of strong \ion{Fe}{2} emission, and the potential \ion{Na}{1} detection suggest a unique chemical composition for the planetary body that is populating the debris disk. It is worth noting however that the white dwarf star is several thousand degrees cooler than those of WD J0347+1624 and WD 1226+110, so the difference could be due to the excitation mechanism for the circumstellar gas. 

\subsection{WD J0644$-$0352 \label{sec:0644}}

\begin{figure}[h!]
\epsscale{1.2}
\plotone{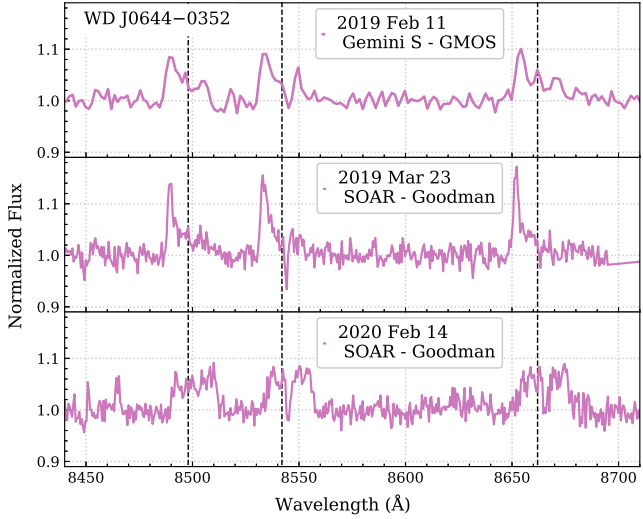}
\caption{The weak, highly asymmetric calcium infrared triplet emission lines of WD J0644$-$0352 were observed transitioning to nearly symmetric state in early 2020. An absorption feature seen near the center of the emission lines is presumed to be atmospheric. \label{fig:0644}}
\end{figure}

\begin{figure}[h!]
\epsscale{1.2}
\plotone{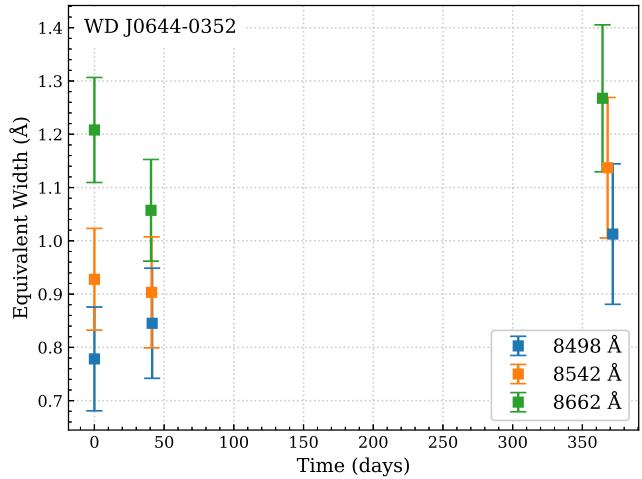}
\caption{No significant changes are seen in the equivalent widths of the calcium triplet emission lines in our follow-up of WD J0644-0352. \label{fig:wd0644eqw}}
\end{figure} 

The calcium infrared triplet lines from the gaseous debris around WD J0644$-$0352 were initially seen as weak and highly asymmetric, with a strong blue-shifted peak and little initial evidence of a red-shifted emission component in February 2019 (see Figure \ref{fig:0644}). The lines appeared almost single-peaked, and the lack of a clear red-shifted component clouded the origin of the emission. Binary systems consisting of a white dwarf star and a highly irradiated stellar or sub-stellar companion are also known to show both an infrared excess and emission lines from the companion (e.g. \citealt{lon17:mnras471}), and their radial velocity variable, single-peaked emission lines can easily be mistaken for an emission feature from an asymmetric debris disk. A recent example of this is SDSS J114404.74+052951.6, initially understood as a gaseous disk around a white dwarf \citep{guo15:apjl810}, but later shown to demonstrate radial velocity variability consistent with chromospheric emission from a heated companion \citep{flo20:aas}.

Such extreme line profile asymmetry has been seen in the gaseous components of white dwarf debris disks, notably at the maximum phases of evolution for HE 1349$-$2305 \citep{den18:apj854}. In cases like this, multi-epoch spectra or evidence of a red-shifted emission shelf is needed to confirm that the emission lines are from a circumstellar disk. Our follow-up in March of 2019 showed little evolution of the profiles, however by February of 2020 the profiles had evolved to a near-symmetric phase with a doppler-broadened structure, confirming the presence of a Keplerian disk and profile evolution similar to HE 1349$-$2305. The timescale of variations indicates a period between four and ten years. No significant variation in the equivalent width of the lines is seen between our three epochs as shown in Figure \ref{fig:wd0644eqw}. 

\subsection{WD J1622+5840 \label{sec:1622}}

\begin{figure}[h!]
\epsscale{1.2}
\plotone{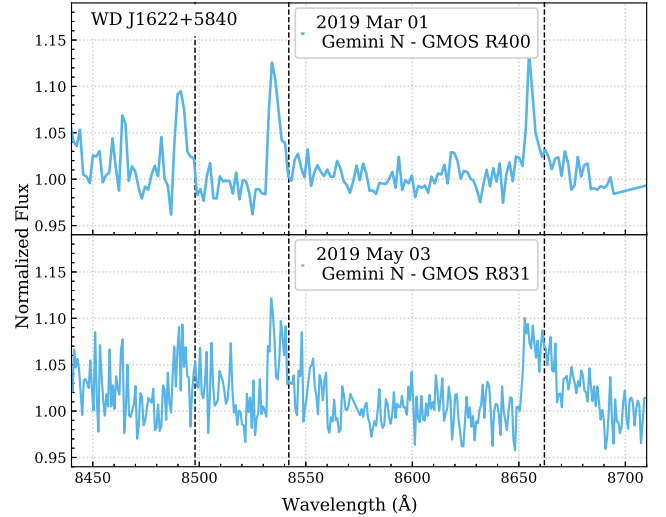}
\caption{Similar to WD J0644$-$0352, the weak and highly asymmetric calcium infrared triplet emission of WD J1622+5840 show evidence of evolution towards a more symmetric profile in our early follow-up. \label{fig:1622}}
\end{figure}

In the initial March 2019 spectrum shown in Figure \ref{fig:1622}, the calcium infrared triplet emission profiles for WD J1622+5840 were remarkably similar to those of WD J0644$-$0352. They appeared nearly single-peaked and completely blue-shifted, prompting the need for higher-resolution and multi-epoch follow-up to confirm their nature. In a second spectrum taken in May 2019 with a higher-resolution setup, we saw that the asymmetric, blue-shifted lines do indeed have a weak, red-shifted shelf that is indicative of an eccentric, Keplerian disk. We have not attained further follow-up to confirm the variability of the emission lines, but the similarities with the gaseous debris observed around WD J0644$-$0352 and HE 1349$-$2305 suggest they should transition to a more symmetric phase over time. 

In addition to the calcium infrared triplet emission, we see evidence of \ion{O}{1} in emission at 7770\AA\ and 8450\AA\ in both the low and high-resolution spectra, shown in Figure \ref{fig:wd1622_em}. The \ion{O}{1} features have similar peak fluxes to the calcium triplet emission lines.

\subsection{WD J2100+2122 \label{sec:2100}}

\begin{figure}[h!]
\epsscale{1.2}
\plotone{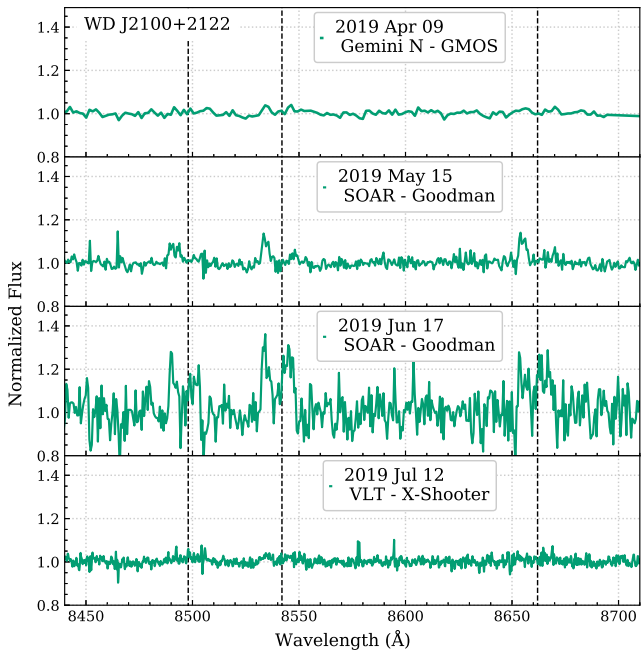}
\caption{A rapid increase in strength is seen in the calcium infrared triplet emission of WD J2100+2122, suggesting a highly dynamic environment with ongoing gas production or excitation. The changes are also seen in the equivalent width measurements presented in Figure \ref{fig:wd2100eqw}. \label{fig:wd2100}}
\end{figure}

\begin{figure}[h!]
\epsscale{1.2}
\plotone{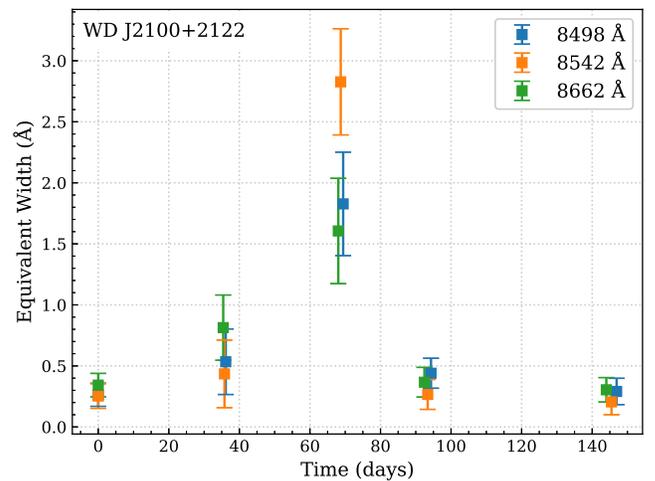}
\caption{The changes in equivalent width of the calcium infrared triplet emission profiles of WD J2100+2122 demonstrate the rapid appearance and disappearance of the emission features over a timescale of a few months. The first four epochs correspond to the four spectra shown in Figure \ref{fig:wd2100}. \label{fig:wd2100eqw}}
\end{figure}

The initial April 2019 spectrum showed a very marginal detection of calcium infrared triplet emission, shown in Figure \ref{fig:wd2100}. Follow-up in May 2019 showed signs of increasing strength in the calcium infrared triplet, with the appearance of an asymmetric, blue-shifted profile. The increase in strength and change in morphology continued in June 2019, where we observed a double-peaked emission profile structure. By July 2019, the emission profiles had decreased in strength, completing a return to the initial state. In Figure \ref{fig:wd2100eqw} we show our measurements of the equivalent widths of the emission lines over time, which show the entire departure and return taking place within two months. 

A similar change in strength of the calcium triplet emission features has been observed in the debris disk surrounding WDJ1617+1620 \citep{wil14:mnras445}, wherein the emission profiles first increased and then slowly declined over the course of ten years, suggesting the fading is related to a return to a nascent state following recent collision or other stochastic event \citep{wil14:mnras445}. The changes in strength we have observed in WD J2100+2122 occur on a timescale of months as opposed to years, increasing and decreasing in strength within sixty days. As we have only been monitoring the system for a few months, it is unlikely that we have managed to catch a one-off brightening event. However with only a single event observed, we can not say whether we expect the event to repeat and on what timescale. Still, cyclical changes in emission strength are an intriguing possibility as the timescale could indicate an ongoing interaction between an existing disk and a larger planetesimal on a wide orbit. Given the dynamic origin of the material that populates white dwarf debris disks, such interactions are expected during the tidal disruption process \citep{per20:mnras493} and have been observed by the evolution of the disrupting, transiting debris \citep{van15:nat526,van19:arxiv}. 

In addition to the calcium infrared triplet emission, a series of \ion{Fe}{2} emission is observed throughout the spectrum and a weak \ion{O}{1} line is detected, as shown in Figure \ref{fig:wd2100_em}. In some cases the \ion{Fe}{2} emission is stronger than the calcium infrared triplet emission. At just over 25,000\,K, the white dwarf star is very hot for hosting gaseous and dusty debris which could be responsible for the high strength of the \ion{Fe}{2} emission with respect to the calcium infrared triplet. Unfortunately, on many epochs we do not have the spectral coverage to search for changes in the strength in the \ion{Fe}{2} emission features to compare with the changes we see in the calcium infrared triplet emission. 

\section{Discussion \label{sec:discussion}}

The discovery of five new white dwarf debris disks with dusty infrared excesses and gaseous debris in emission provides a substantial increase of the known sample, increasing the number from seven to 12 \citep{man16:mnras462}, including the hottest white dwarf discovered to host such a disk, WD J2100+2122. The increase is expected given the influx of known white dwarf stars from \emph{Gaia}, and the continued spectroscopic follow-up of white dwarfs with infrared excesses and large samples of stars expected to be observed as part of upcoming spectroscopic surveys are certain to discover more \citep{man20:mnras493, mel20:arxiv}.

With targeted photometric follow-up, we confirm the \emph{WISE} infrared excess reported for all five systems. Using a single-temperature blackbody fit to the infrared excess, we conclude that the infrared excess is likely due to dust, though we are not able to constrain the dust properties without further modeling. The infrared excess seen around WD J0611$-$6931 is exceedingly strong for interpretation as a flat, dusty debris disk, making it a good target for detailed studies of alternative debris disk geometries \citep{jur07:aj133}. 

WD J0347+1624 and WD J2100+2122 show decreases between the \emph{WISE} catalog photometry and our \emph{Spitzer} follow-up, and the lack of nearby sources in the corresponding \emph{Spitzer} images suggests that the \emph{WISE} photometry is unlikely to be contaminated, and that intrinsic variability may be responsible for the difference. This is particularly interesting in the context of the recent findings of \citet{swa20:mnras496}, that shows dusty white dwarfs with gaseous debris in emission tend to show stronger infrared variations than those without gaseous debris in emission. Both targets are sufficiently bright for detection in the NEOWISE-R Single Exposure Source Table\footnote{Available at \url{https://irsa.ipac.caltech.edu}}, and should be monitored over time for continued variation. 

\begin{figure*}[ht!]
\epsscale{1.15}
\plotone{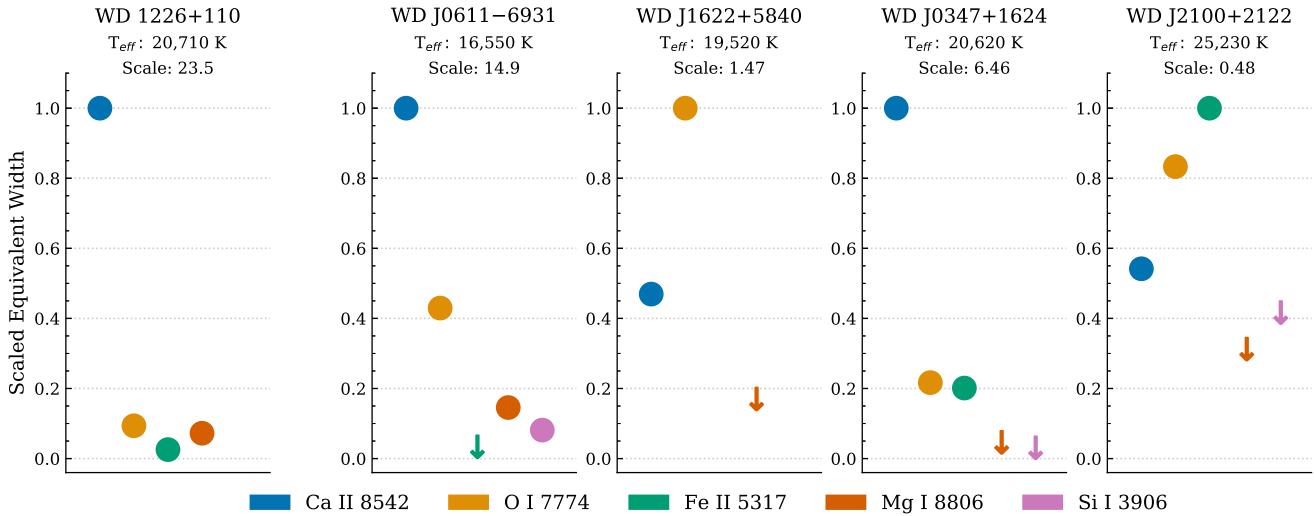}
\caption{A visual representation of the relative strengths of the new emission species detected in four of the systems, compared against the prototypical system WD 1226+110 on the left \citep{man16:mnras455}. Detections are shown as circles and upper limits as downward facing arrows, and the points have been scaled by the maximum equivalent width detected in each spectrum. WD J1622+5840 and WD J2100+2122 in particular stand out as having emission lines from additional species that are stronger than the calcium infrared triplet, though the measurements for WD J2100+2122 were taken after the peak in calcium triplet emission strength. The comparison demonstrates the diversity of circumstellar environments these systems exhibit. \label{fig:alleqws}}
\end{figure*}

All five of the white dwarf debris disk systems show signs of active accretion via the detection of atmospheric calcium. Some show absorption features from other metal species, such as Mg, Si, and Fe, and are suitable for detailed abundance analyses to explore the chemical compositions of the accreted bodies and accretion rates of the material to compare with the greater sample (e.g. \citealt{xu19:aj158, swa19:mnras490}). These analyses would benefit from a more accurate determination of the white dwarf atmospheric properties using spectroscopic methods, which we leave for future works. The possibility of magnetism in WD J1622+5840 has strong implications for its circumstellar environment and should be further investigated. Spectropolarimetric measurements could quickly determine if the white dwarf is indeed weakly magnetic \citep{lan19:aap628}, and spectra of nearby stars could be obtained to rule out contributions from interstellar absorption.

Three of the systems we present show significant evolution of their emission line profiles over their first year of observation. With these new discoveries, it is clear there are at least two classes of long-term variability exhibited by white dwarfs with gaseous debris that are not necessarily mutually exclusive. WD J0347+1624 and WD J0644$-$0352 display the asymmetric profile evolution that has been associated with precessing eccentric debris around SDSS J0845+2257, WD 1226+110, HE 1349-2305, and WD 1145+017 \citep{wil15:mnras451,man16:mnras455,den18:apj854,cau18:apjl852,mir18:apj857}, while WD J2100+2122 exhibits both morphological variability and stochastic variations that had previously only been observed in SDSS J1617+1620 \citep{wil14:mnras445}. The short variability timescales for the emission lines in WD J0347+1624, WD J0644$-$0352, and WD J2100+2122 make them attractive targets for continued follow-up. 

The detection of multiple strong metal species in emission for several of the systems also provides future works the opportunity for direct compositional analysis of the gaseous debris. Previously, additional metal species were detected in only two of the seven similar systems known to host dusty and gaseous debris in emission, including \ion{Fe}{2} in Ton 345 \citep{mel10:apj722} and \ion{Fe}{2}, \ion{O}{1}, \ion{Mg}{1}, and \ion{Mg}{2} in WD 1226+110 \citep{mel10:apj722, har11:aap530, man16:mnras455}. In contrast, WD J0611$-$6931, WD J1622+5840, WD J2100+2122 all show additional features which are comparable in strength or stronger than the calcium infrared triplet, which is visualized in Figure \ref{fig:alleqws}. Note that the absolute strengths of the emission lines also span nearly two orders of magnitude, as denoted by the scale factors used to normalize the axes. The detections of multiple strong species in emission and significant variability in the strength of the calcium infrared triplet in WD J2100+2122 emphasize that future searches for these systems should rely on broad-band and multi-epoch spectroscopy when available. 

The range of emission strengths demonstrate that these systems host a diverse range of circumstellar environments. The different excitation species could be signs of unique planetary compositions, or more complex environmental factors such as the debris disk geometry or heating mechanisms. Recent advances in the modeling of the gaseous environments around white dwarf stars show that with enough detections of individual excitation species, these degeneracies can be lifted \citep{gan19:nat576,for20:apj888}. In addition to the continued variability monitoring, the direct compositional analyses of the debris in the systems presented here is an exciting prospect for future studies. 

\acknowledgments

We would like to thank A. Nitta, S. Kleinman, and C. Melis for useful discussions throughout the course of this work. We are also grateful to the anonymous referee for providing a detailed review that helped improve this manscript. AB acknowledges funding from a Royal Society Dorothy Hodgkin Fellowship. This work is also partly supported by the Heising-Simons Foundation via the 2019 Scialog program on Time Domain Astrophysics. 

The authors wish to recognize and acknowledge the very significant cultural role and reverence that the summit of Maunakea has always had within the indigenous Hawaiian community.  We are most fortunate to have the opportunity to conduct observations from this mountain.

Based on observations obtained at the international Gemini Observatory, a program of NSF's OIR Lab and processed using the Gemini IRAF package and DRAGONS (Data Reduction for Astronomy from Gemini Observatory North and South, which is managed by the Association of Universities for Research in Astronomy (AURA) under a cooperative agreement with the National Science Foundation on behalf of the Gemini Observatory partnership: the National Science Foundation (United States), National Research Council (Canada), Agencia Nacional de Investigaci\'{o}n y Desarrollo (Chile), Ministerio de Ciencia, Tecnolog\'{i}a e Innovaci\'{o}n (Argentina), Minist\'{e}rio da Ci\^{e}ncia, Tecnologia, Inova\c{c}\~{o}es e Comunica\c{c}\~{o}es (Brazil), and Korea Astronomy and Space Science Institute (Republic of Korea). Based on observations collected at the European Southern Observatory under ESO programmes 0103.C-0431(B), 1103.D-0763(D), and 0104.C-0107(A). Some of the data presented herein were obtained at the W. M. Keck Observatory, which is operated as a scientific partnership among the California Institute of Technology, the University of California and the National Aeronautics and Space Administration. The Observatory was made possible by the generous financial support of the W. M. Keck Foundation. Some of the data is based on observations obtained at the Southern Astrophysical Research (SOAR) telescope, which is a joint project of the Minist\'{e}rio da Ci\^{e}ncia, Tecnologia, Inova\c{c}\~{o}es e Comunica\c{c}\~{o}es (MCTIC) do Brasil, the U.S. National Optical Astronomy Observatory (NOAO), the University of North Carolina at Chapel Hill (UNC), and Michigan State University (MSU).

This publication makes use of data products from the Two Micron All Sky Survey, which is a joint project of the University of Massachusetts and the Infrared Processing and Analysis Center/California Institute of Technology, funded by the National Aeronautics and Space Administration and the National Science Foundation. This research has made use of the NASA/ IPAC Infrared Science Archive, which is operated by the Jet Propulsion Laboratory, California Institute of Technology, under contract with the National Aeronautics and Space Administration. This publication makes use of data products from the Wide-field Infrared Survey Explorer, which is a joint project of the University of California, Los Angeles, and the Jet Propulsion Laboratory/California Institute of Technology, and NEOWISE, which is a project of the Jet Propulsion Laboratory/California Institute of Technology. WISE and NEOWISE are funded by the National Aeronautics and Space Administration. This work has made use of data from the European Space Agency (ESA) mission {\it Gaia} (\url{https://www.cosmos.esa.int/gaia}), processed by the {\it Gaia} Data Processing and Analysis Consortium (DPAC, \url{https://www.cosmos.esa.int/web/gaia/dpac/consortium}). Funding for the DPAC has been provided by national institutions, in particular the institutions participating in the {\it Gaia} Multilateral Agreement.\\

\vspace{5mm}
\facilities{Spitzer (IRAC), WISE, Gaia, Keck:I (HIRES), Gemini:South (GMOS-S, Flamingos-2), Gemini:North (GMOS-N, NIRI), SOAR (Goodman), VLT:Kueyen (X-Shooter)}
\software{astropy \citep{ast13:aap558}, DRAGONS \citep{lab19:aspc523}}

\appendix
\restartappendixnumbering 

\section{Additional Emission Lines Detected and Observation Logs}

In this appendix we present figures showing the additional emission species detected in addition to our spectroscopic and photometric observation logs. The additional emission detections were done by eye, and may not consitute an exhaustive list of all lines present in the spectra. Wavelengths for the lines are taken from the NIST Atomic Spectral Database and referenced in air.

\begin{figure*}[ht!]
\gridline{\fig{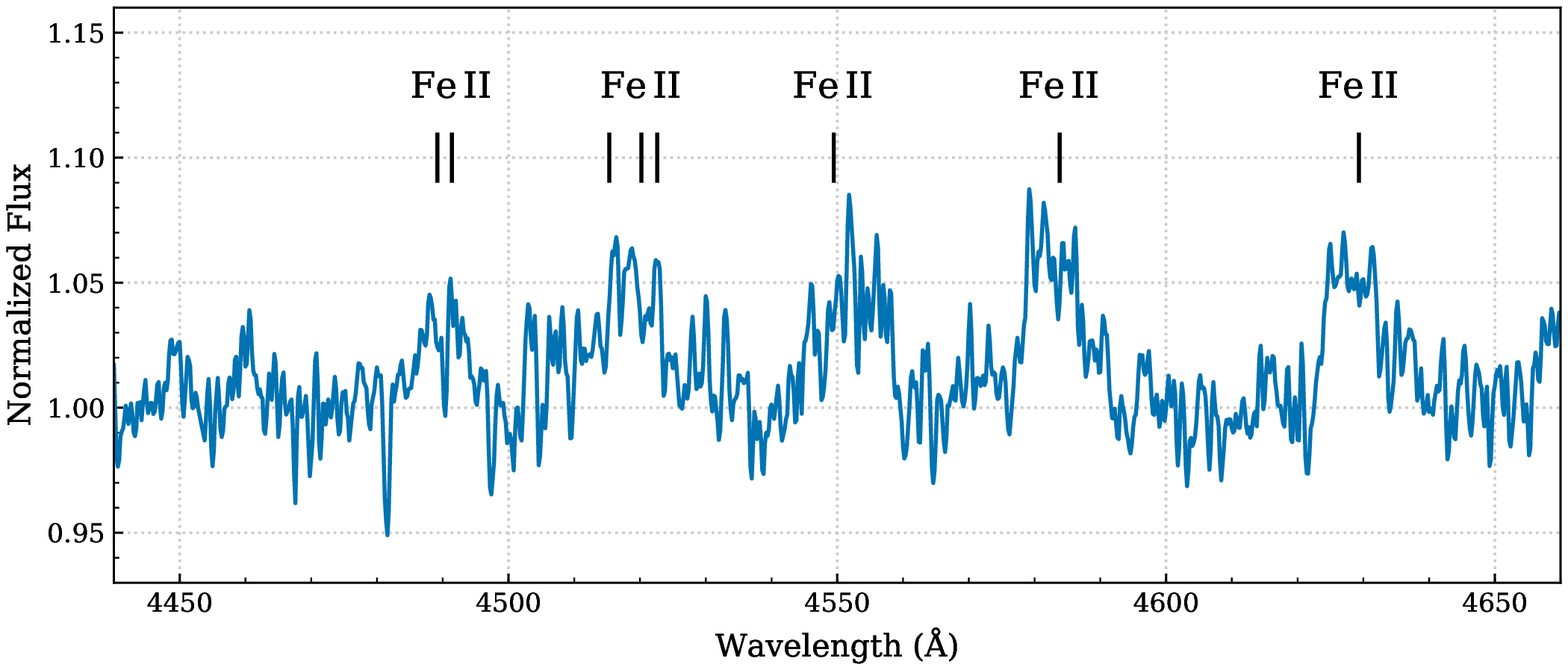}{0.49\textwidth}{ }\fig{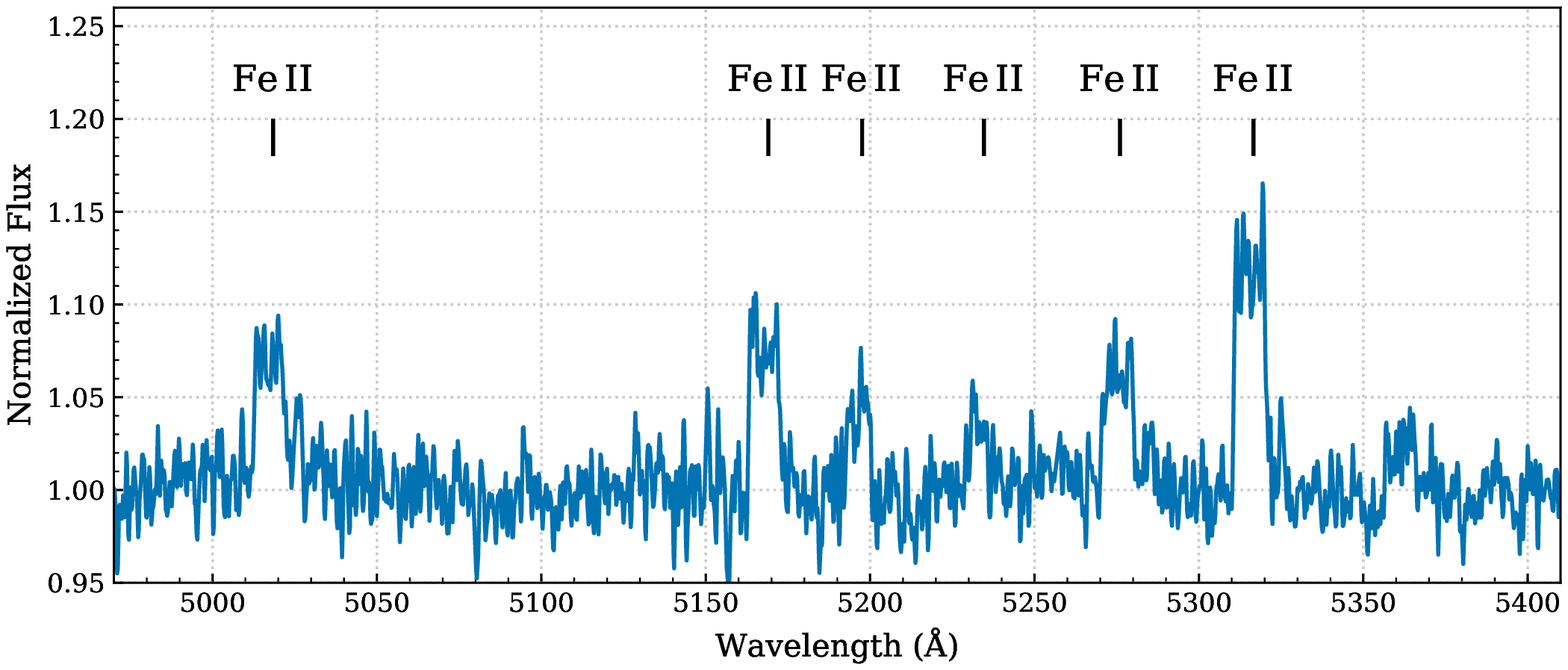}{0.49\textwidth}{}}
\vspace{-0.8cm}
\gridline{\fig{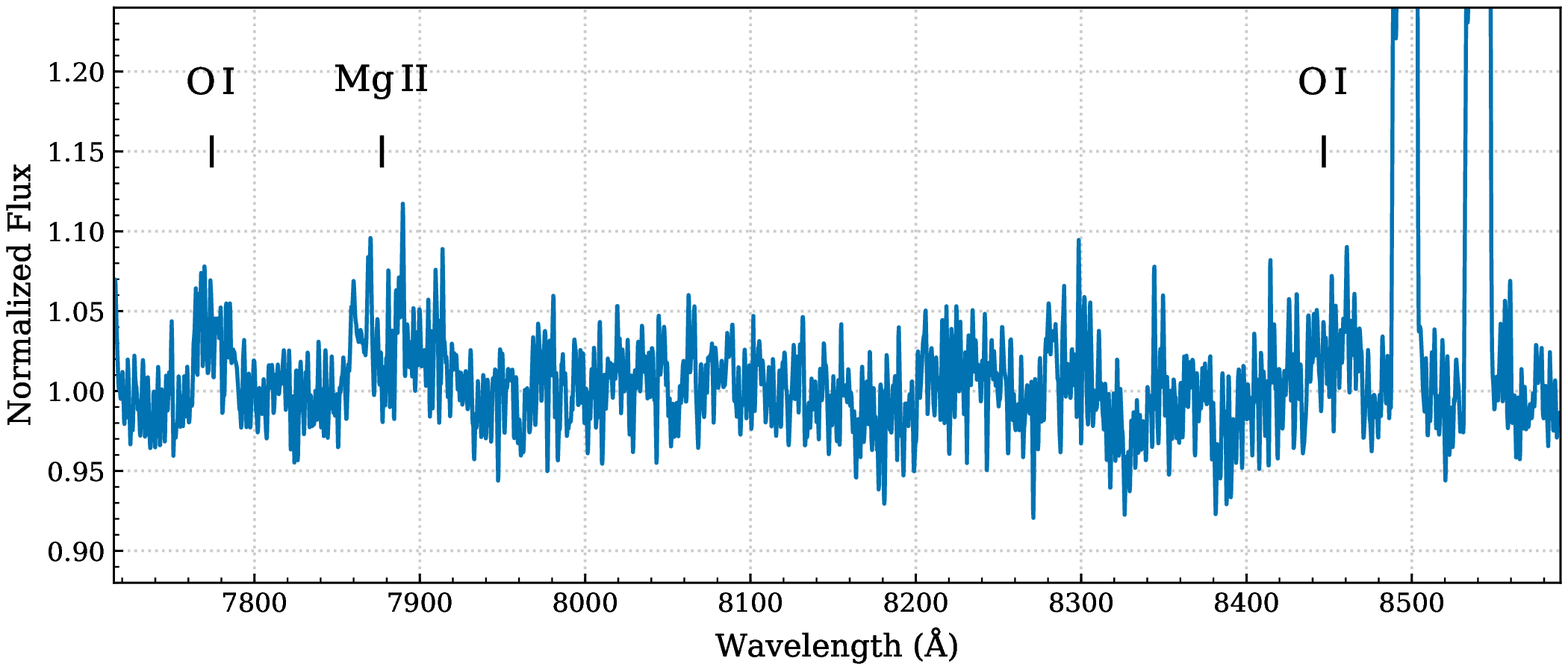}{0.49\textwidth}{}}
\caption{VLT/X-Shooter spectra of WD J0347+1625 shows several metal species in emission in addition to the calcium infrared triplet. \ion{Fe}{2} species from the 3d$^6$(5D)4p upper energy level are detected at 4489.18, 4491.40, 4515.33, 4520.22, 4522.62, 4549.46, 4583.82, 4629.33, 5018.44, 5169.03, 5197.57, 5234.62, 5276.00, and 5316.61 \AA. \ion{O}{1} species from the 2s$^2$2p$^3$($^4$S$^0$)3p upper energy level are detected in a blend around 7774.17 and 8446.36 \AA\, in the gaseous debris of WD J0347+1624. A weak, blended line of \ion{O}{1} and \ion{Mg}{2} may also be present near 7890 \AA. \label{fig:wd0347_em}}
\end{figure*}

\begin{figure*}[ht!]
\gridline{\fig{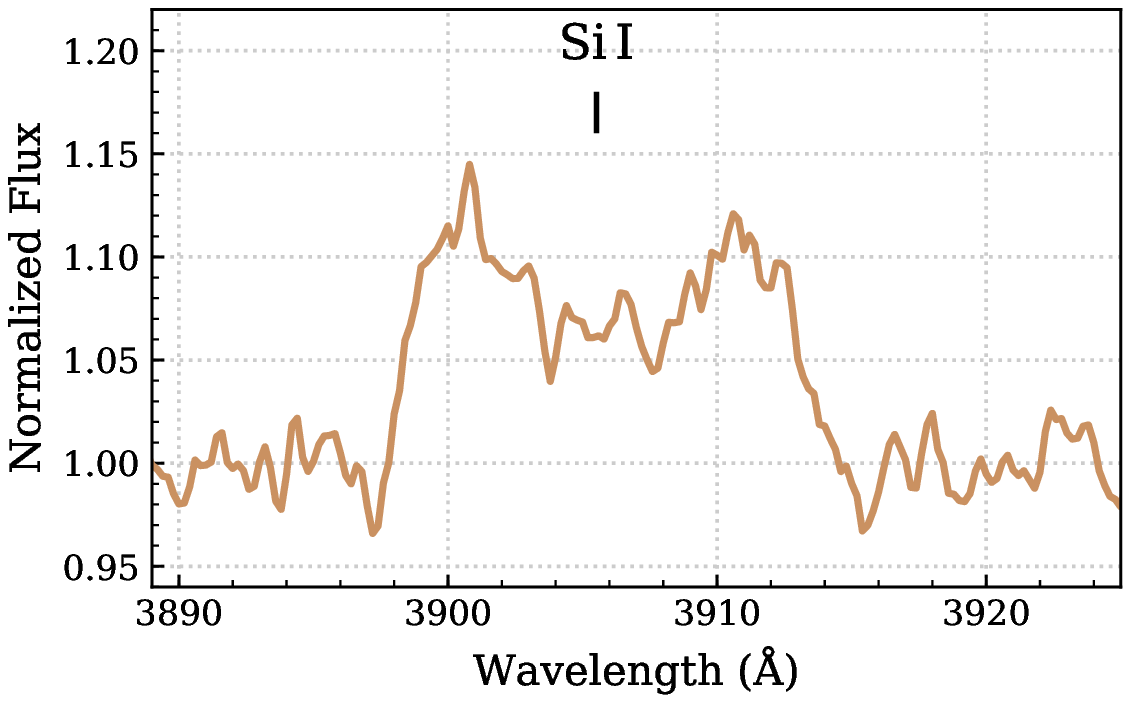}{0.45\textwidth}{}\fig{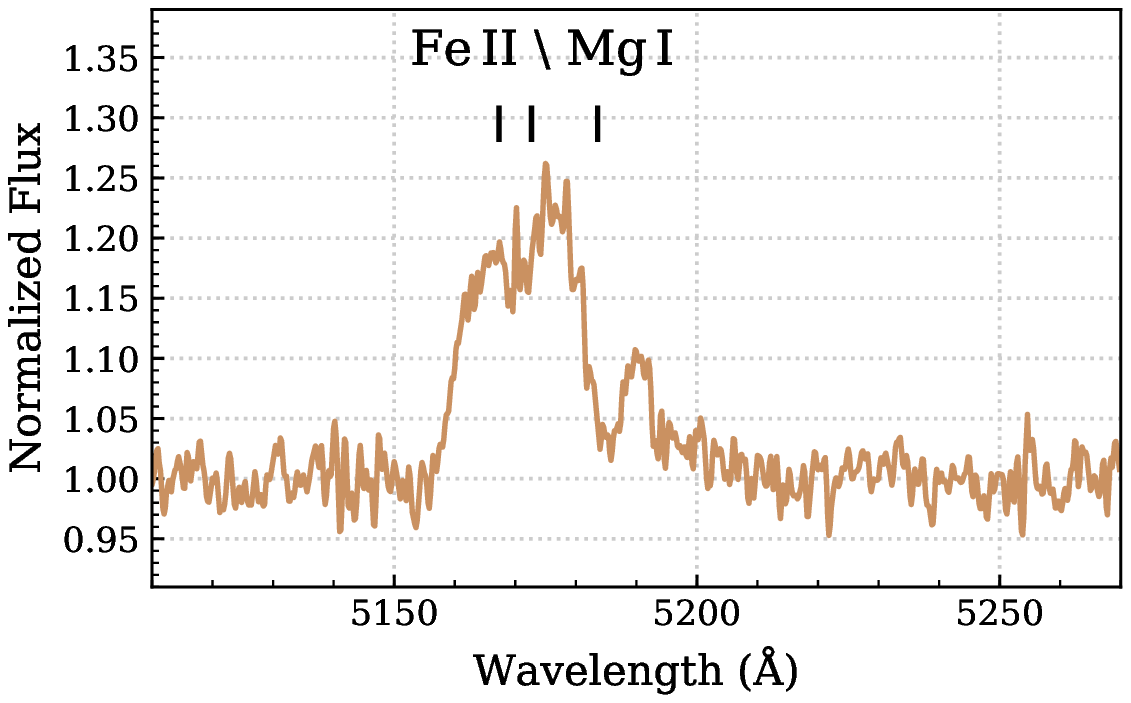}{0.45\textwidth}{ }}
\vspace{-0.8cm}
\gridline{\fig{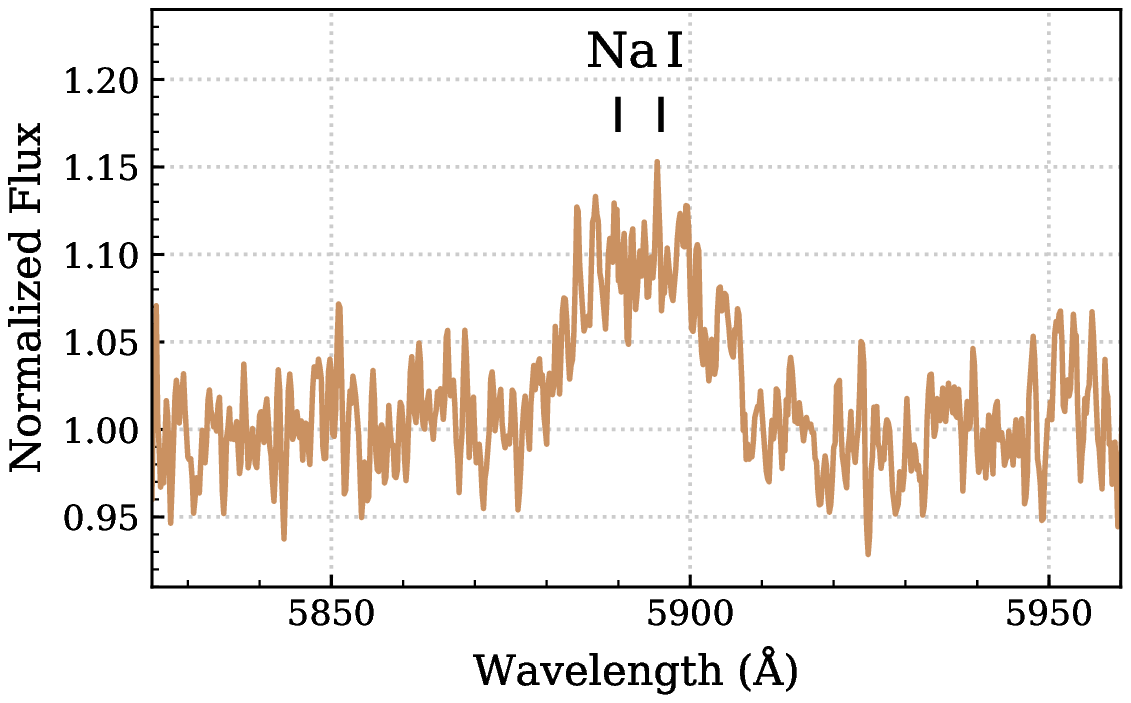}{0.45\textwidth}{}\fig{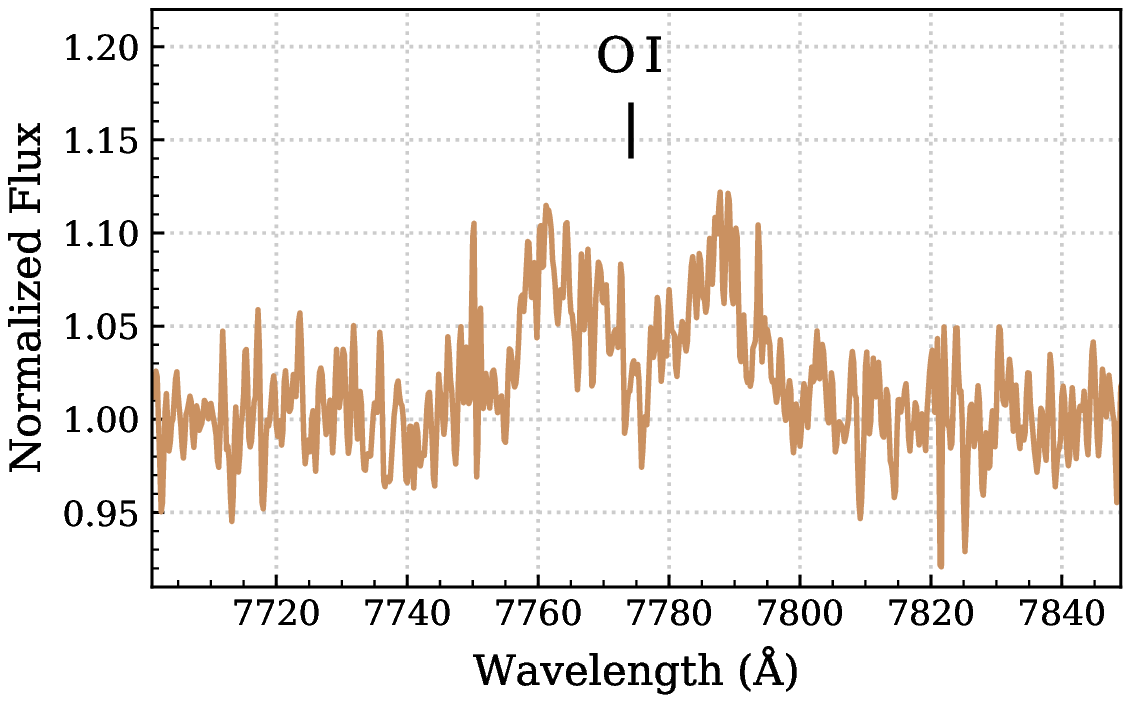}{0.45\textwidth}{}}
\caption{Additional metal species seen in emission in the VLT/X-Shooter spectra of WD J0611$-$6931. A blend of \ion{Mg}{1} species from the 3s4s upper energy level are detected at 5167.32, 5172.68, 5183.60 \AA\, and potentially \ion{Fe}{2} from the 3d$^6$(5D)4p upper energy level at 5169.03 \AA. No other strong \ion{Fe}{2} lines are detected. \ion{Si}{1} from the 3s$^2$3p4s upper energy level is detected at 3905.52 \AA, and \ion{O}{1} species from a blend around 7774.17 \AA\, are detected in emission from the gaseous debris of WD J0611$-$6931. The \ion{O}{1} and \ion{Si}{1} features share a similar profile shape and width in velocity space as the calcium infrared triplet features, suggesting they are coming from the same emitting region. Finally, an emission feature is detected near the \ion{Na}{1} doublet at 5889.95 and 5895.92 \AA. \label{fig:wd0611_em}}
\end{figure*}

\begin{figure*}[ht!]
\gridline{\fig{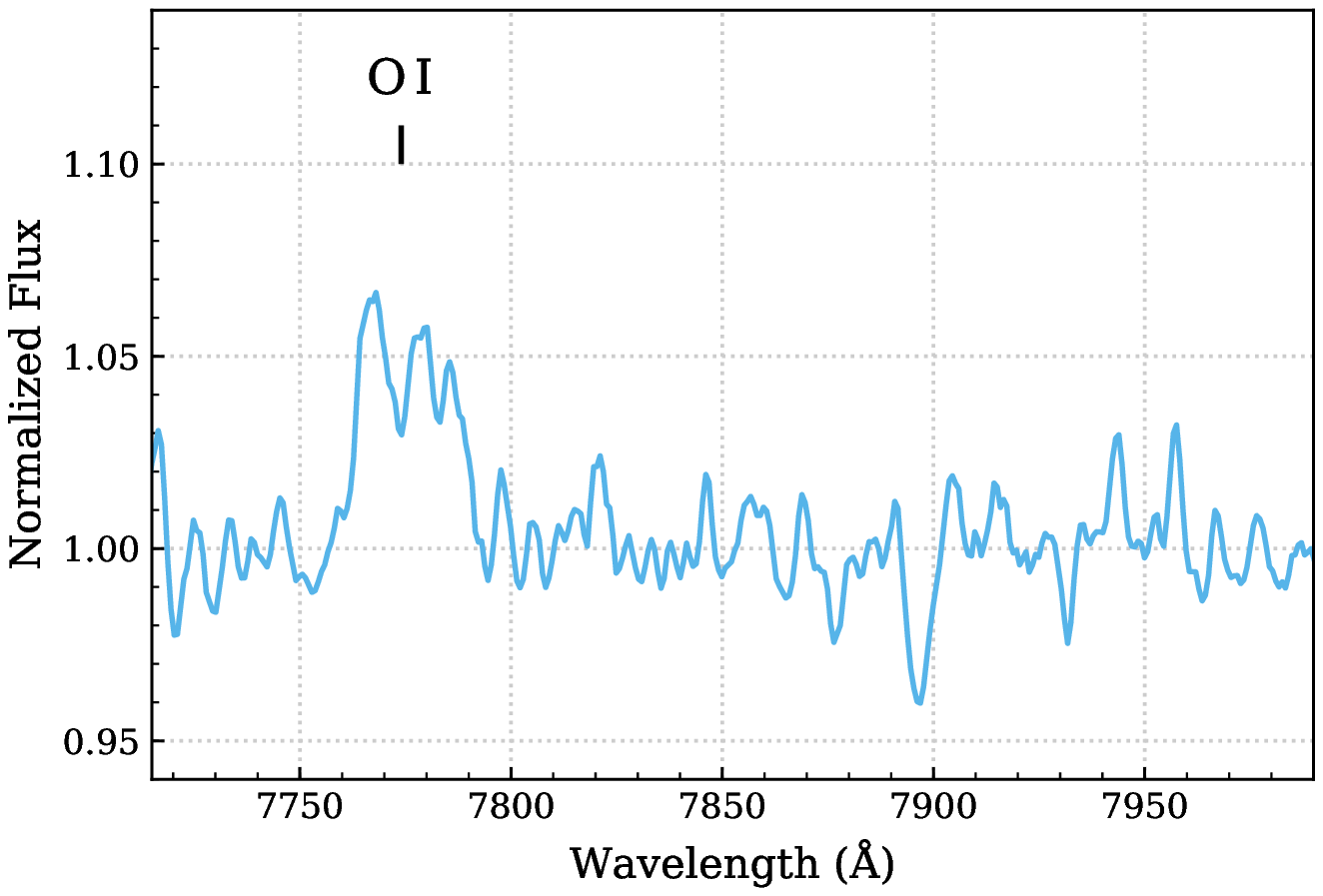}{0.45\textwidth}{ }\fig{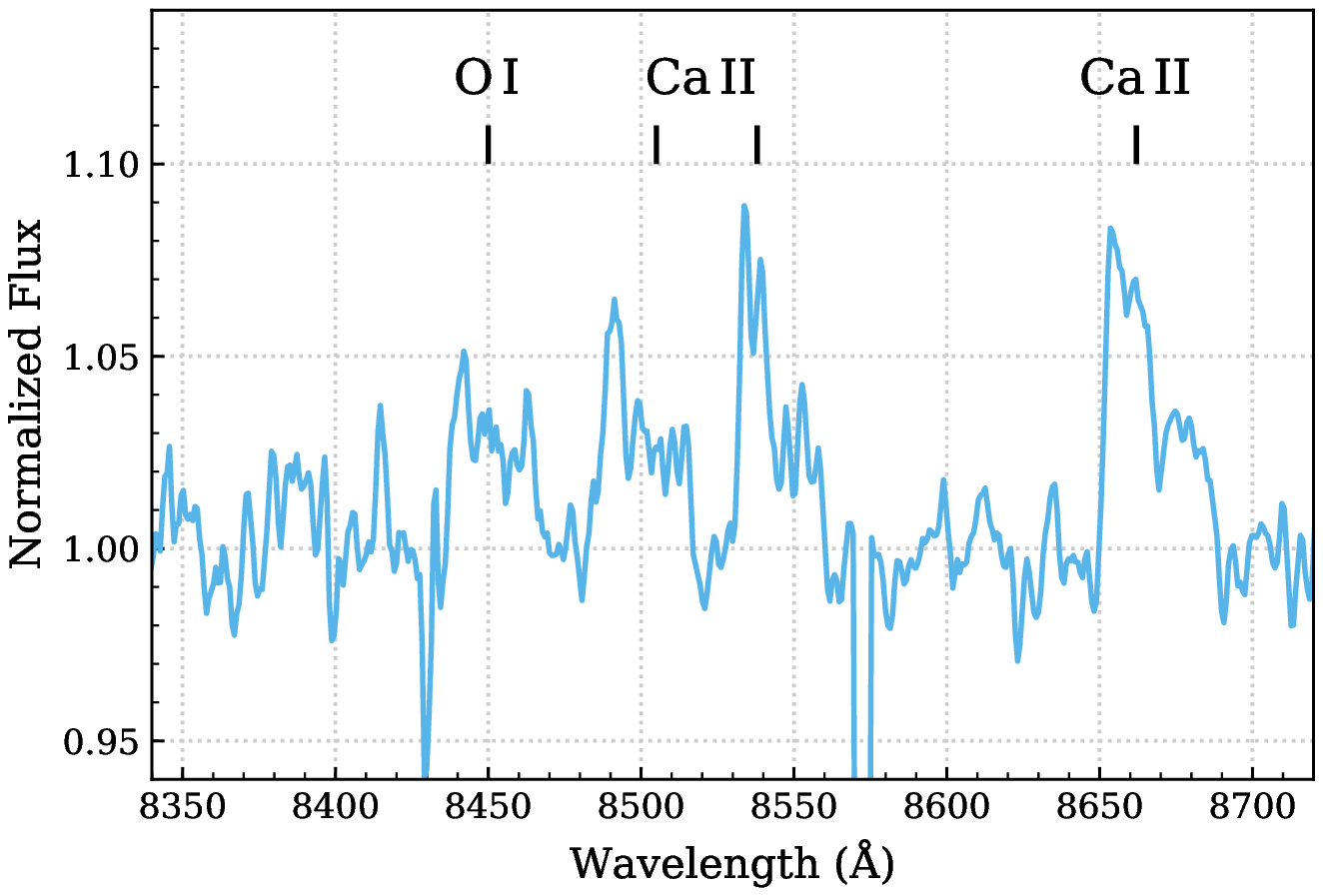}{0.45\textwidth}{}}
\caption{\ion{O}{1} species are seen in emission in the smoothed Gemini-N/GMOS spectra of WD J1622+5840. \ion{O}{1} species from the 2s$^2$2p$^3$($^4$S$^0$)3p upper energy level are detected in a blend around 7774.17 and 8446.36 \AA\, in the gaseous debris of WD J1622+5840, comparable in strength to the calcium infrared triplet. \label{fig:wd1622_em}}
\end{figure*}

\begin{figure*}[ht!]
\gridline{\fig{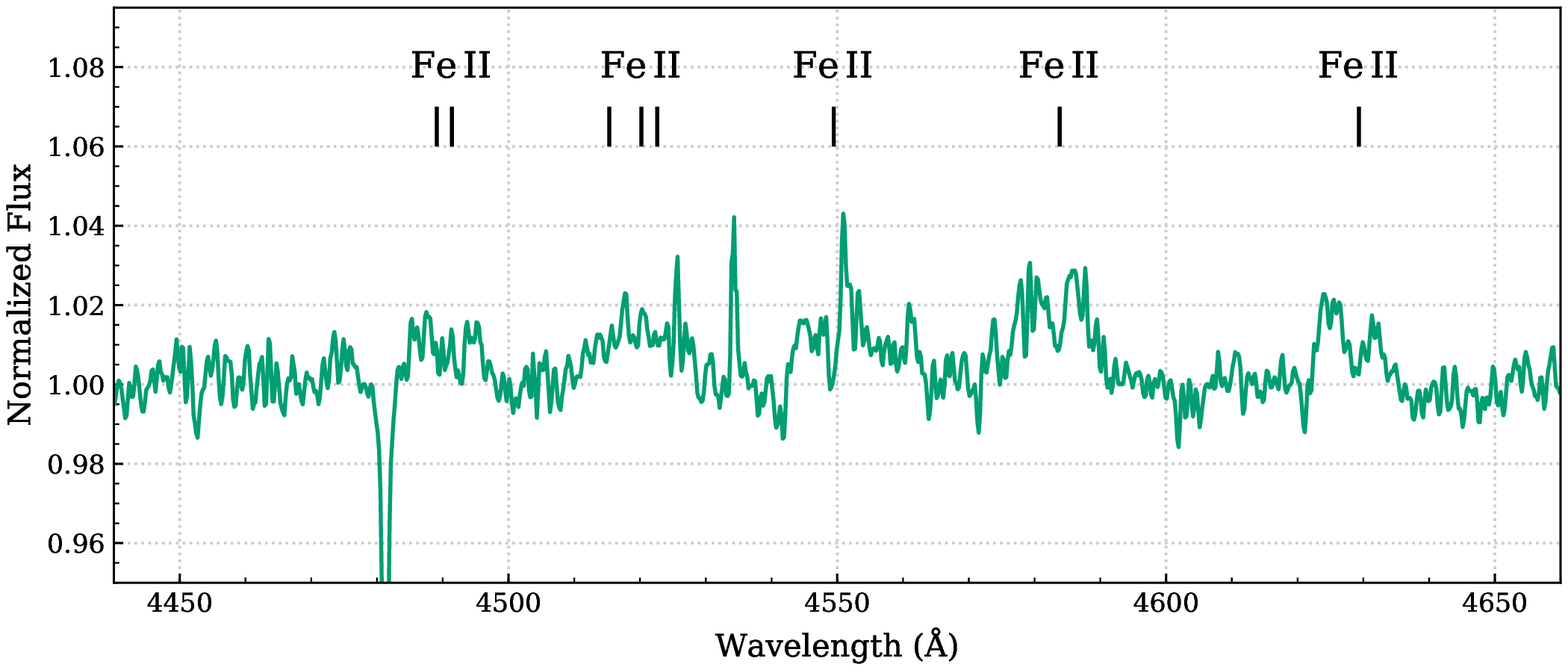}{0.49\textwidth}{ }\fig{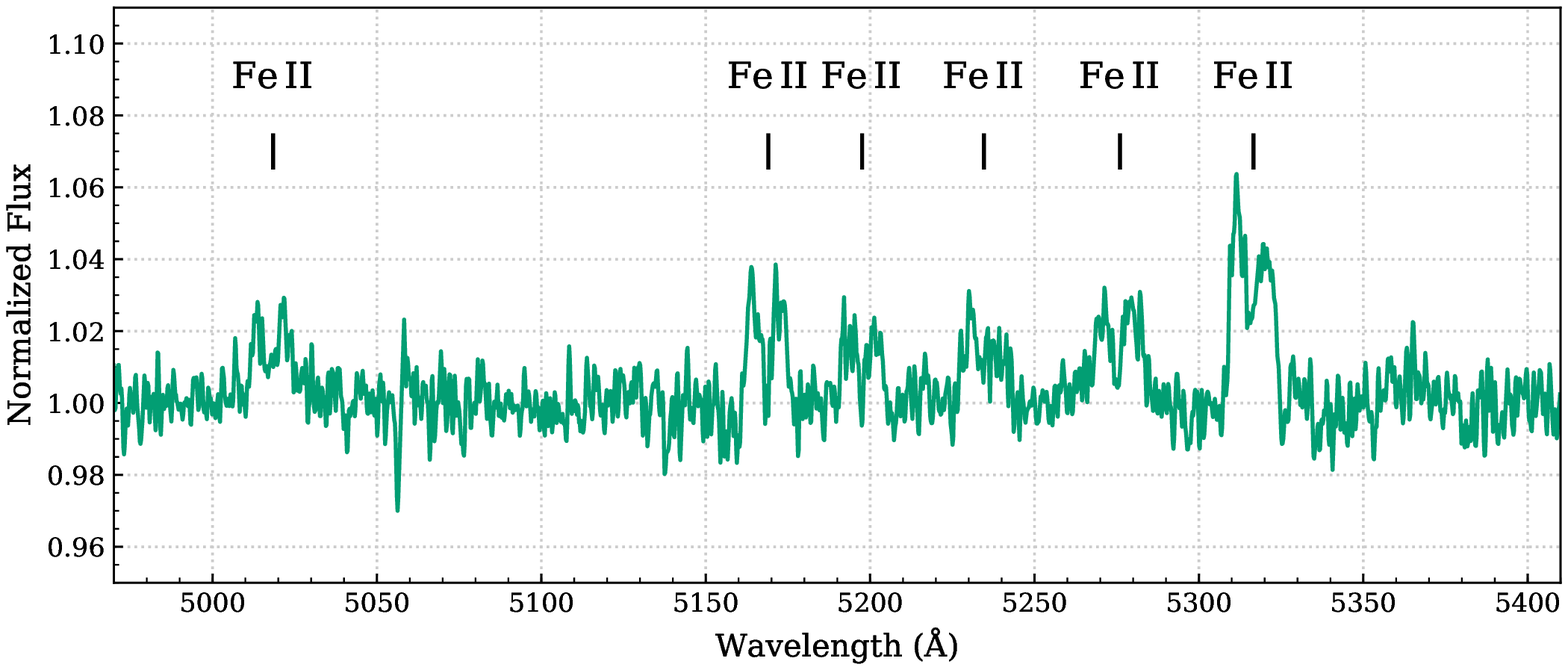}{0.49\textwidth}{}}
\vspace{-0.8cm}
\gridline{\fig{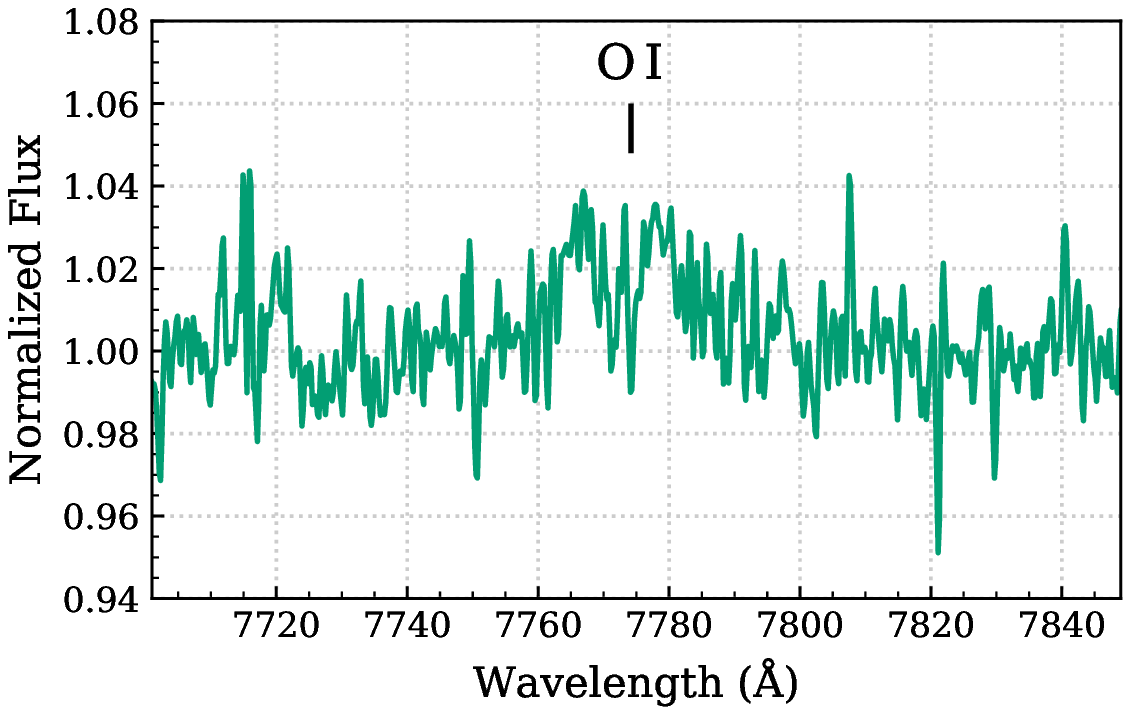}{0.4\textwidth}{}}
\caption{Several iron species are seen in emission in the VLT/X-Shooter spectrum of WD J2100+2122. \ion{Fe}{2} species from the 3d$^6$(5D)4p upper energy level are detected at 4489.18, 4491.40, 4515.33, 4520.22, 4522.62, 4549.46, 4583.82, 4629.33, 5018.44, 5169.03, 5197.57, 5234.62, 5276.00, and 5316.61 \AA\, in the gaseous debris of WD J2100+2122.  A weak \ion{O}{1} species from the 2s$^2$2p$^3$($^4$S$^0$)3p upper energy level is detected in a blend around 7774.17 \AA. The iron lines are comparable in strength to the calcium infrared triplet. \label{fig:wd2100_em}}
\end{figure*}

\begin{deluxetable*}{lllllll}[ht!]
\tablecaption{Observation log for collected spectroscopic data. \label{tab:obsspec}}
\tablehead{\colhead{Target} & \colhead{Telescope/Instrument} & \colhead{Program ID} & \colhead{UT Date} & \colhead{Coverage (\AA)} & \colhead{Resolving Power} & \colhead{Integration Time (s)}}
\startdata
WD J0347+1624  & Gemini-N/GMOS-N & GN-2019A-FT-202 & 2019 Jan 24 & 6200 $-$ 10500 & 1900 & 5x300 \\
 & Keck/HIRESb & 2019B\_N072 & 2019 Dec 05 & 3130 $-$ 5960 & 37,000 & 3x1800 \\
 & SOAR/Goodman & Partner-UNC & 2018 Dec 03 & 7550 $-$ 8720 & 3000 & 5x600 \\
 & SOAR/Goodman & Partner-UNC & 2019 Nov 14 & 7550 $-$ 8720 & 3000 & 9x600 \\
 & VLT/X-Shooter & 1103.D-0763(D) & 2019 Dec 05 & 3000 $-$ 10400 & 5100/8800 & 4x400\\
WD J0611$-$6931 & VLT/X-Shooter & 0104.C-0107(A) & 2019 Oct 15 &  3000 $-$ 10400 & 5100/8800 & 2x1700 \\
WD J0644$-$0352 & Gemini-S/GMOS-S & GS-2019A-FT-201 & 2019 Feb 11 & 6200 $-$ 10500 & 1900 & 5x300  \\
 & SOAR/Goodman & Partner-UNC & 2019 Mar 23 & 7550 $-$ 8720 & 3000 & 12x600 \\
 & SOAR/Goodman & Partner-UNC & 2020 Feb 14 & 7550 $-$ 8720 & 3000 & 16x600 \\
 & VLT/X-Shooter & 103.C-0431(B) & 2019 Sep 14 & 3000 $-$ 10400 & 5100/8800 & 2x1700 \\
WD J1622+5840 & Gemini-N/GMOS-N & GN-2019A-FT-202 & 2019 Mar 01 & 6200 $-$ 10500 & 1900 & 5x300 \\
& Gemini-N/GMOS-N & GN-2019A-FT-209 & 2019 May 03 & 7300 $-$ 9600 & 3800 & 7x300 \\
 & Keck/HIRESb & 2019B\_N072 & 2019 Jul 10 & 3130 $-$ 5960 & 37,000 & 3x1800 \\
WD J2100+2122 & Gemini-N/GMOS-N & GN-2019A-FT-202 & 2019 Apr 09 & 6200 $-$ 10500 & 1900 & 4x300 \\
  & Keck/HIRESb & 2019B\_N072 & 2019 Jul 10 & 3130 $-$ 5960 & 37,000 & 3x1200 \\
   & SOAR/Goodman & Partner-UNC & 2019 May 15 & 7550 $-$ 8720 & 3000 & 5x600 \\
 & SOAR/Goodman & Partner-UNC & 2019 Jun 17 & 7550 $-$ 8720 & 3000 & 6x600 \\
  & SOAR/Goodman & Partner-UNC & 2019 Sep 02 & 7550 $-$ 8720 & 3000 & 6x600 \\
  & VLT/X-Shooter & 0103.C-0431(B) & 2019 Jul 12 & 3000 $-$ 10400 & 5100/8800 & 4x400\\
\enddata
\end{deluxetable*}

\begin{deluxetable*}{llllll}[ht!]
\tablecaption{Observation log for collected photometric data. \label{tab:obsphot}}
\tablehead{\colhead{Target} & \colhead{Telescope/Instrument} & \colhead{Program ID} & \colhead{Date} & \colhead{Filters (Integration Time)}}
\startdata
WD J0347+1624  &  Gemini-S/Flamingos-2  &  GS-2019B-FT-204 & 2018 Oct 17  & \emph{J} (6x10s), \emph{H} (19x6s), \emph{Ks} (19x10s)\\ 
  &  \emph{Spitzer}/IRAC  & 14220 & 2019 Jun 02  & Ch$\,$1 (11x30s), Ch$\,$2 (11x30s)\\ 
WD J0611$-$6931  &  Gemini-S/Flamingos-2  &  GS-2018B-Q-404 & 2018 Dec 22  & \emph{J} (6x15s), \emph{H} (29x6s) \\ 
  &  Gemini-S/Flamingos-2  &  GS-2018B-Q-404 & 2018 Dec 30  & \emph{Ks} (17x20s)\\ 
  &  \emph{Spitzer}/IRAC  &  70062 & 2010 Dec 22  & Ch$\,$1 (5x30s) \\ 
WD 00644$-$0352 &  Gemini-S/Flamingos-2  &  GS-2019B-FT-204 & 2018 Oct 17  & \emph{J} (6x10s), \emph{H} (19x6s) \\ 
  &  Gemini-S/Flamingos-2  &  GS-2019B-FT-204 & 2018 Oct 20  & \emph{Ks} (19x10s) \\ 
  &  \emph{Spitzer}/IRAC  &  14220 & 2019 Jul 05  & Ch$\,$1 (11x30s), Ch$\,$2 (11x30s)\\ 
WD J1622+5840  &  Gemini-N/NIRI  &  GN-2019A-Q-303 & 2019 Jun 15  & \emph{J} (10x10s), \emph{H} (26x10s), \emph{K} (26x10s)\\ 
  &  \emph{Spitzer}/IRAC  &  14220 & 2019 Jun 08  & Ch$\,$1 (11x30s), Ch$\,$2 (11x30s)\\ 
WD J2100+2122  &  Gemini-N/NIRI  &  GN-2019A-Q-303 & 2019 Jun 30  & \emph{J} (10x10s), \emph{H} (26x10s), \emph{K} (26x10s)\\ 
  &  \emph{Spitzer}/IRAC  &  14220 & 2019 Sep 09  & Ch$\,$1 (11x30s), Ch$\,$2 (11x30s)\\ 
\enddata
\end{deluxetable*}

\clearpage

\bibliography{wdexoplanets}
\bibliographystyle{aasjournal}

\end{CJK}
\end{document}